\documentclass[prd,amsmath,superscriptaddress,twocolumn]{revtex4}

\usepackage{graphicx}
\usepackage{amsmath}
\usepackage{amssymb}
\usepackage{graphics}
\usepackage{hyperref}
\usepackage{mathrsfs}
\usepackage{bm}
\usepackage[usenames,dvipsnames]{color}

\newcommand{\mpcinvtxt}{$\mathrm{Mpc}^{-1}$}
\newcommand{\mpcinv}{\mathrm{Mpc}^{-1}}
\newcommand{\ain}{a_\mathrm{in}}
\newcommand{\camb}{{\tt{CAMB}}}
\newcommand{\cmbfast}{{\tt{CMBFAST}}}
\newcommand{\copter}{{\tt{Copter}}}
\newcommand{\corrfour}[4]{\left<\varphi_{#1}\varphi_{#2}\varphi_{#3}\varphi_{#4}\right>}
\newcommand{\corrthree}[3]{\left<\varphi_{#1}\varphi_{#2}\varphi_{#3}\right>}
\newcommand{\corrtwo}[2]{\left<\varphi_{#1}\varphi_{#2}\right>}
\newcommand{\delcb}{\delta_\mathrm{cb}}
\newcommand{\dirac}{\delta_\mathrm{D}}
\newcommand{\dlin}{\delta_\mathrm{lin}}
\newcommand{\enw}{\epsilon}
\newcommand{\fcb}{f_\mathrm{cb}}
\newcommand{\fnu}{f_\nu}
\newcommand{\nnw}{\mu}
\newcommand{\Omegam}{\Omega_\mathrm{m}}
\newcommand{\Pcb}{P_\mathrm{cb}}
\newcommand{\Plin}{P_\mathrm{lin}}
\newcommand{\Pnw}{P_\mathrm{nw}}
\newcommand{\Ponethree}{P^{(1,3)}}
\newcommand{\Ptwotwo}{P^{(2,2)}}
\newcommand{\rhode}{\rho_\mathrm{de}}
\newcommand{\rhom}{\rho_\mathrm{m}}
\newcommand{\rhombar}{\bar\rho_\mathrm{m}}

\newcommand{\Tabg}{\colorbox{lightgray}}

\definecolor{lightgray}{gray}{0.9}

\begin{document}

\title{Large-Scale Structure Formation with Massive Neutrinos and 
Dynamical Dark Energy} 

\author{Amol Upadhye}
\address{High Energy Physics Division, Argonne National
  Laboratory, 9700 S. Cass Ave., Lemont, IL 60439} 
\address{Kavli Institute for Cosmological Physics, The University
  of Chicago, Chicago, IL 60637}
\author{Rahul Biswas}
\author{Adrian Pope}
\address{High Energy Physics Division, Argonne National
  Laboratory, 9700 S. Cass Ave., Lemont, IL 60439} 
\author{Katrin Heitmann}
\author{Salman Habib}
\address{High Energy Physics Division, Argonne National
  Laboratory, 9700 S. Cass Ave., Lemont, IL 60439}  
\address{Kavli Institute for Cosmological Physics, The University
  of Chicago, Chicago, IL 60637}
\address{Mathematics and Computer Science Division, Argonne
  National Laboratory, 9700 S. Cass Ave., Lemont IL 60439}
\author{Hal Finkel}
\address{ALCF, Argonne National Laboratory, 9700 S. Cass Ave.,
  Lemont, IL 60439}%
\author{Nicholas Frontiere}
\address{High Energy Physics Division, Argonne National
  Laboratory, 9700 S. Cass Ave., Lemont, IL 60439} 
\address{Department of Physics, The University of Chicago,
  Chicago, IL 60637} 

\date{\today}

\begin{abstract}
  Over the next decade, cosmological measurements of the large-scale
  structure of the Universe will be sensitive to the combined effects
  of dynamical dark energy and massive neutrinos. The matter power
  spectrum is a key repository of this information. We extend
  higher-order perturbative methods for computing the power spectrum
  to investigate these effects over quasi-linear scales. Through
  comparison with N-body simulations we establish the regime of
  validity of a Time-Renormalization Group (Time-RG) perturbative
  treatment that includes dynamical dark energy and massive
  neutrinos. We also quantify the accuracy of Standard (SPT),
  Renormalized (RPT) and Lagrangian Resummation (LPT) perturbation
  theories without massive neutrinos. We find that an approximation
  that neglects neutrino clustering as a source for nonlinear matter 
  clustering predicts the Baryon Acoustic Oscillation (BAO) peak position to
  $0.25\%$ accuracy for redshifts $1 \leq z \leq 3$, justifying the
  use of LPT for BAO reconstruction in
  upcoming surveys. We release a modified version of the public
  {\copter} code which includes the additional physics discussed in the
  paper.

\end{abstract}
\maketitle

\section{Introduction}
\label{sec:introduction}
\subsection{Dynamical Dark Energy and Neutrinos} 
\label{subsec:de_nu}
The original discovery of the late-time acceleration of the
Universe~\cite{Riess_etal_1998,Perlmutter_etal_1999} has been
confirmed by multiple cosmological probes over the last fifteen years.
The underlying cause of this acceleration, as well as its connection
to fundamental physics, however, remains to be clarified. Although the
cosmological ``Standard Model'' provides an excellent description of
the latest data~\cite{Vikhlinin:2008ym, Conley_etal_2011,
  Suzuki_etal_2012, Kilbinger:2012qz, Hinshaw:2012aka, Hou:2012xq,
  Calabrese:2013jyk, Ade_etal_2013xvi, Anderson_etal_2013}, the
cosmological constant is beset with problems of extreme fine-tuning,
at least given our current level of theoretical understanding (for
reviews see, Refs.~\cite{Weinberg_rmp, Nobbenhuis_2006,
  Polchinski:2006gy, Caldwell:2009,Martin:2012bt}). For this reason,
it is extremely important to constrain the evolution of the dark
energy equation of state, and to see if the associated equation of
state parameter, $w(z)$, deviates from the constant value, $w=-1$,
characteristic of a cosmological constant. While phenomenological
models for $w(z)$ can be postulated for comparison to observed
data, one class of solutions known as early dark energy allows for the
energy density responsible for the acceleration to be much larger at
earlier times, with possible connections to fundamental
physics~\cite{Binetruy:1999, Zlatev:1999, Steinhardt_etal_1999,
  Armendariz-Picon_etal_2000, Barriero00, Bueno_Sanchez_2006,
  Alam:2010, Alam:2011}.

Investigations of dynamical dark energy are conducted using two
distinct types of observational probes: {\em (i)} constraints on the
homogeneous expansion of the Universe, and {\em (ii)} the growth of
large-scale structure, driven primarily by the gravitational dynamics
of cold dark matter (CDM). Via consistency relations, these results
can also be used to study the validity of general relativity in
describing the dynamics of the Universe (see, e.g.,
Ref.~\cite{ishak06}), although this is not our concern here.

Besides dark energy, a major contribution of modern cosmology to
fundamental physics lies in constraining the sum of neutrino masses,
as well as the number of neutrino species.  Massive neutrinos act as a
radiation component in the early universe, but as a warm dark matter
fluid at late times.  The high velocity of neutrinos makes them
difficult to bind gravitationally, suppressing the growth of structure
in a scale-dependent manner.  Aside from being interesting in their
own right, massive neutrinos can be degenerate with the dark energy
density and the equation of state~\cite{hannestad05}, which
necessitates an analysis including both effects. Therefore, it is
essential for future precision measurements to include neutrinos as a
component of the analysis, along with a time-varying dark energy
equation of state or a modified theory of gravity. A thorough
investigation of this issue within the Fisher matrix formalism is
presented in Ref.~\cite{font-ribera13} for spectroscopic redshift
surveys such as BOSS (Baryon Oscillation Spectroscopic
Survey~\cite{Schlegel:2009hj}) or DESI (Dark Energy Spectroscopic
Instrument~\cite{Levi:2013gra}), but also including input from the
cosmic microwave background (CMB), in particular,
Planck~\cite{Ade_etal_2013xvi}, and weak gravitational lensing
surveys such as DES (Dark Energy Survey~\cite{Abbott:2005bi}) and LSST
(Large Synoptic Survey Telescope~\cite{Abell:2009aa}). A similar
analysis for surveys like Euclid~\cite{Refregier:2010ss} can be found
in Ref.~\cite{Basse13}. The power of purely large-scale measurements,
such as baryon acoustic oscillations (BAO) is significantly reduced by
the uncertainty in neutrino masses, and employing broadband galaxy
power at much smaller scales becomes important in improving our
ability to extract information about dark energy as well as the
neutrino sector~\cite{font-ribera13}.

Due to their large thermal velocities, $v_{{\rm th}}(z)$, neutrinos do
not cluster at scales smaller than the free-streaming scale
$k_{\rm{FS}}(z) \sim H (z)/v_{\rm th}(z)$. For neutrinos turning
non-relativistic in the matter dominated regime, the comoving
free-streaming scale has a maximum value at the time when the
neutrinos become relativistic. Thus, at length scales larger than
those set by this maximum, $k_{\rm nr}$, neutrinos cluster in the same
way as dark matter, while at smaller scales their contribution to
clustering is much smaller, leading to a suppression of the total
matter power spectrum. In linear perturbation theory this suppression
increases with increasing wave-number asymptoting to a value of $\sim
8 \Omega_\nu/\Omega_{\rm m}.$ Since this effect can be observed at
length scales too small for linear perturbation theory to hold, it is
essential to compute the nonlinear matter power spectrum~\cite{Wong_2008}.

The effect of neutrinos at sufficiently large length scales can be
studied using perturbation theory; at smaller length scales, matter
clustering treated via N-body methods can be used to extend the
predictive reach. Moreover, in view of the above discussion, this has
to be done in the presence of a varying dark energy equation of
state. Our purpose here is to present both perturbative and N-body
results for the matter fluctuation power spectrum in the presence of
neutrinos and dynamical dark energy. (We do not consider the case of
modified gravity here.)

Our results are useful in multiple ways. First, they provide reliable
predictions for the matter power spectrum on large scales for ongoing
and upcoming BAO measurements like BOSS~\cite{Schlegel:2009hj},
DESI~\cite{Levi:2013gra}, and CHIME (the Canadian Hydrogen Intensity
Mapping Experiment~\cite{chime:web}), in particular at higher
redshifts.  Second, in order to build prediction tools for power
spectra well into the nonlinear regime, perturbation theory is very
useful to anchor the predictions at high accuracy on large
scales. Previously, based on a finite number of cosmological models,
we have produced emulators for the power spectrum enabling fast
parameter estimation of $w$CDM cosmologies~\cite{Lawrence09,
  Heitmann13}. However, ongoing and future surveys will attempt to go
beyond $w$CDM to constrain both neutrino masses and a time dependent
equation of state of dark energy. In preparation for such surveys
(see, e.g. Ref.~\cite{desc12}), we use the HACC (Hardware/Hybrid
Accelerated Cosmology Code) N-body framework~\cite{Habib09, Pope10,
  Habib12}, extended to include both a time varying dark energy
equation of state and an approximate treatment of massive neutrinos,
to compute the matter power spectrum. Third, the accurate treatment of
neutrinos in N-body simulations is non-trivial as discussed in more
detail below. We study the validity of different higher order
perturbation theory implementations for a $\Lambda$CDM cosmology for
which we have high-accuracy simulations. We then use these results to
gauge the inaccuracies induced by an approximate treatment of
neutrinos in simulations.

Dynamical dark energy can be treated in two different ways. One can
either begin with a model, specified by an action, and aim to
constrain its parameters.  For example, a scalar field quintessence
model~\cite{Peebles88,Ratra88} can be written down with a power law
potential whose parameters can then be determined by the data.  Such
models can be subdivided, for example, into ``freezing'' and
``thawing'' classes, in which the scalar field moves respectively
toward or away from a stationary point in its
potential~\cite{Caldwell05}.  Canonical scalar fields have
pressure-to-energy-density ratios (``equations of state'') ranging
between $-1$ and $1$, and their sound speeds are equal to the speed of
light; however, k-essence models relax both of these
restrictions~\cite{Armendariz-Picon01}.

The second, more phenomenological, approach to dark energy is to
parameterize its equation of state and sound speed as functions of
time or scale factor.  Although an action is necessary for predicting
the effects of dark energy across a range of energy and distance
scales, a dark energy which does not cluster gravitationally and does
not couple to any other particle can really only be constrained on
cosmological scales, so such a parameterization is sufficient.  Here
we adopt the commonly-used form~\cite{Chevalier01, Linder03},
\begin{equation}
w(a) = w_0 + w_a(1-a),
\label{e:cpl}
\end{equation}
and we assume a sound speed $c_s^2=1$.  This representation of $w(a)$
smoothly parameterizes a large range of models including freezing and
thawing scalar fields, phantom energy with $w<-1$, and early dark
energy.  One of its limitations is that constraints on early dark
energy imply $w_0+w_a \lesssim 0$, limiting the rate at which $w(a)$
can change at recent times~\cite{Upadhye05}.  However, since the data
are not powerful enough to constrain a large number of dark energy
parameters, this parameterization is a reasonable compromise, and is
used in analyzing the results from many surveys.

\subsection{Perturbation Theory and N-Body Simulations}

Over the past several years, higher-order cosmological perturbation
theory has been crafted into a useful tool for understanding the
growth of large-scale structure.  Applying the continuity and Euler
equations to an effective matter fluid with an irrotational velocity
field, perturbation theory lets us predict the power spectrum of
large-scale structure at early times and at moderately nonlinear
scales.  Although the scale-dependent growth rate in massive neutrino
models is incompatible with most perturbative methods, so-called
``Time Renormalization Group'' (Time-RG) perturbation theory
accommodates massive neutrinos by directly integrating the evolution
equations for the power spectrum~\cite{Pietroni08, Lesgourgues09}.  In
this work we extend the publicly available \copter~perturbation theory
code~\cite{Carlson09, COPTER} to include massive neutrinos and
dynamical dark energy.  In our perturbative treatment, as well as in
our computation of the scale-dependent growth, we treat neutrinos as a
linear source for growth of cold dark matter and baryonic density
perturbations.

The regime of validity of perturbation theory cannot be calculated
rigorously, and at sufficiently small scales the fluid approximation
breaks down~\cite{Pueblas_etal_2009, Valageas_2011, Anselmi12,
  Pietroni_etal_2012}.  Currently the most reliable test of
perturbation theory is a direct comparison with N-body simulations,
which approximate dark matter as a collection of point particles,
obtaining dynamical Monte Carlo solutions to the Vlasov-Poisson system
of equations. Such a comparison is given in Ref.~\cite{Carlson09}
using a large range of perturbation theory methods for $\Lambda$CDM
cosmologies, and quantifying the accuracy of each perturbative method.
Here we use N-body simulations to test Time-RG and a few other
perturbation theories, for models with dynamical dark energy and
massive neutrinos, up to redshifts $z=3$.
 
While N-body simulations are the accepted way to compute the nonlinear
matter power spectrum of cold dark matter, including neutrinos in the
simulation as particles is difficult because of their large thermal
velocities, and artificial clustering induced by having multiple
species of particles with very different particle
masses. Consequently, different groups have computed nonlinear
corrections to the matter power spectrum due to massive neutrinos by
adopting different approaches. Perturbation theory was extended to the
quasi-nonlinear regime by using standard second order perturbation
theory in Ref.~\cite{Saito08}.  In Ref.~\cite{Agarwal11}, the authors
account for neutrinos in the initial conditions for the CDM particles
in the simulations, and add the linearly evolved neutrino fluctuations
to the particle fluctuations to obtain the total power spectrum,
ignoring the nonlinear interaction of the neutrinos with the dark
matter. An alternative approach to studying the effect of neutrinos at
small scales by using the halo model is pursued in
Ref.~\cite{Abazajian05}.

Gravitational interactions between neutrinos and dark matter particles
were self-consistently incorporated in Ref.~\cite{Brandbyge08} by
including neutrino particles in N-body simulations with a thermal
velocity sampled from the appropriate Fermi-Dirac distribution in
addition to a flow velocity to set up initial conditions, typically
starting the (neutrino) simulation at very late times so that the
neutrino thermal velocity is relatively small
~\cite{Brandbyge08,Viel10,Bird12}. The aim of this was to avoid
difficulties of the sort encountered in Ref.~\cite{gardini99} and in
other earlier simulation efforts.

To extend the treatment further, in Refs.~\cite{Viel10,Brandbyge09},
the authors propagate a linearly evolved neutrino perturbation on a
grid, and CDM particles in an N-body simulation that evolves under the
influence of self-gravity as well as the potential sourced by the
neutrino fluid on the grid. Further improvements to this method were
introduced in Ref.~\cite{Brandbyge10} where this approach was combined
with a particle representation to study sub-Mpc effects of neutrino
clustering, and in Ref.~\cite{Ali-Hamoud13} who also evolve the
neutrino density in the nonlinear potentials sourced by the CDM
particles. These nonlinear corrections show that the nonlinear matter
power spectrum is suppressed to a maximum of $\sim 10
\Omega_\nu/\Omega_{\rm m},$ (as compared to $\sim 8 \Omega_\nu/\Omega_{\rm m}$ 
found in linear theory mentioned earlier) and the suppression decreases for
wavenumbers larger than a certain value $k_{\rm turn}$ which depends
on the neutrino mass.

Our goal is to calculate the power spectrum in the presence of
{\emph{(i)}} massive neutrinos and {\emph{(ii)}} time-varying dark
energy equations of state, only up to $k \approx 0.3~h \mpcinv$,
allowing us to simplify the treatment of neutrinos considerably.  We
include these two new physical ingredients in both simulations and
higher-order perturbation theory, finding that the two methods agree
at the $2\%$ level up to $k\sim 0.1~h\mpcinv$ at $z=0$, and better at
higher redshifts. Through perturbative arguments we show that our
simple approximation for neutrinos in the N-body simulations,
following Refs.~\cite{Saito08, Agarwal11}, is valid at the $1\%$ level
over the entire region of applicability of perturbation theory,
justifying its use in our code.  Finally, we apply our perturbative
calculations to determine the effect on the power spectrum of varying
neutrino masses and the two dark energy equation of state parameters,
finding, for example, that neutrino inhomogeneities have little effect
on the BAO scale.

The remainder of this paper is organized as follows.
Section~\ref{sec:higher-order_perturbation_theory} summarizes
perturbation theory results, incuding Time-RG when massive neutrinos
are included.  Our N-body simulations are described in
Sec.~\ref{sec:n-body_simulations}.  Tests of perturbation theory with
dynamical dark energy and massive neutrinos are conducted in
Sec.~\ref{sec:testing_perturbation_theory} and presented along with
discussions of observable effects. The final results are summarized in
Sec.~\ref{sec:conclusion}.

\section{Higher-order perturbation theory}
\label{sec:higher-order_perturbation_theory}

\subsection{Standard Perturbation Theory}

Consider a universe containing nonrelativistic, non-interacting matter
with density, $\rhom(\vec x,t)$, as well as non-clustering dark energy
with density $\rhode(t)$ and equation of state parameter $w$.
Under the assumptions of an irrotational
velocity field ($\nabla \times \vec v = 0$) and no shell crossings, the matter
can be described in terms of a density contrast, $\delta(\vec x,t) =
(\rhom-\rhombar)/\rhombar$, and a velocity divergence, $\theta(\vec x,
t) = \nabla \cdot \vec v$.  In Fourier space, the continuity and Euler
equations imply
\begin{eqnarray}
\frac{\partial \delta(\vec k,a)}{\partial \log a}
&=&
-\frac{\theta(\vec k,a)}{aH}
\label{e:continuity}
\\
&-&
\int \frac{d^3p \, d^3q}{aH(2\pi)^3} 
\dirac(\vec k \! - \!  \vec p \! - \! \vec q) 
\frac{\vec k \cdot \vec p}{p^2}
\theta(\vec p,a) \delta(\vec q,a)
\nonumber\\
\frac{\partial \theta(\vec k,a)}{\partial \log a}
&=&
-\theta(\vec k,a)
- \frac{3}{2} \Omegam a H \delta(\vec k,a)
\label{e:euler}
\\
&-&
\int \frac{d^3p \, d^3q}{aH(2\pi)^3}
\dirac(\vec k \! - \! \vec p \! - \! \vec q) 
\frac{k^2(\vec p \cdot \vec q)}{2p^2q^2}
\theta(\vec p,a) \theta(\vec q,a) 
\nonumber
\end{eqnarray}
where $a$ is the scale factor, defined to be unity today; $H$ is the
Hubble parameter; and $\dirac$ is the Dirac delta function.  The
second term on the right hand side of Eq.~(\ref{e:euler}) describes
gravitational clustering according to the Poisson equation.  The final
term on the right in each of the above makes the evolution nonlinear.
If these two terms are neglected, Eqs.~(\ref{e:continuity}-\ref{e:euler})
can be integrated easily.  The resulting linear theory describes the
universe on the largest scales.  The $\vec k$-independent linear
growth factor $D(a)$ is the growing-mode solution $\dlin(\vec k,a)$ to
these linearized equations, also normalized to unity today.

The nonlinear evolution equations Eqs.~(\ref{e:continuity}-\ref{e:euler})
can be expressed in a more compact notation.  Define the perturbation
variables, $\varphi_a$, the evolution function matrix, $\Omega_{ab}$,
and the vertex functions, $\gamma_{abc}$, as
\begin{eqnarray}
\varphi_0(\vec k, a) 
&=& 
\delta(\vec k,a) \ain/a
\label{e:varphi_0}
\\
\varphi_1(\vec k,a) 
&=& 
-\theta(\vec k,a) \ain / (a^2 H)
\label{e:varphi_1}
\\
\Omega_{00} 
&=&
-\Omega_{01}
= 1
\label{e:Omega00}
\\
\Omega_{10}(a)
&=&
-\frac{3 \Omegam H_0^2}{2 a^3 H^2}
=
-\frac{3}{2} \Omegam(a)
\label{e:Omega10}
\\
\Omega_{11}(a)
&=&
3 + \frac{d\log H}{d\log a}
\label{e:Omega11}
\\
\gamma_{010}(\vec k, \vec p, \vec q)
&=&
\gamma_{001}(\vec k,\vec q,\vec p)
\label{e:gamma010}
\nonumber\\
&=&
\dirac(\vec k + \vec p + \vec q) 
(\vec p + \vec q)\cdot \vec p / (2 p^2)
\\
\gamma_{111}(\vec k, \vec p, \vec q)
&=&
\dirac(\vec k + \vec p + \vec q) 
(\vec p + \vec q)^2 \vec p \cdot \vec q / (2 p^2 q^2)
\label{e:gamma111}
\end{eqnarray}
where all other $\gamma_{abc}$ are zero, and $\ain \ll 1$ is the
initial value of the scale factor, which we assume to be small enough
that the evolution is linear.  Then the evolution equations
(\ref{e:continuity}-\ref{e:euler}) can be written in the form,
\begin{eqnarray}
\frac{\partial \varphi_a(\vec k,a)}{\partial \log a}
&=&
-\Omega_{ab}(a) \varphi_b(\vec k,a)
\label{e:eom_compact}
\\
&+& 
\!
\frac{a}{\ain} \!\! 
\int \! \frac{d^3p \, d^3q}{(2\pi)^3}
\gamma_{abc}(\vec k, -\vec p, -\vec q) \varphi_b(\vec p,a) \varphi_c(\vec q,a),
\nonumber
\end{eqnarray}
where repeated indices indicate summation.

A thorough description of Standard Perturbation Theory (SPT) is
provided in Ref.~\cite{Carlson09}, which we summarize here.  Let us
begin by assuming an Einstein-de Sitter (EdS) universe, in which
$\Omegam=1$ and there are no species other than cold matter.
Equations~(\ref{e:Omega00}-\ref{e:Omega11}) imply
\begin{equation}
{\mathbf \Omega}
=
\left[
  \begin{array}{cc}
    1            & -1  \\
    -\frac{3}{2} & \frac{3}{2}
  \end{array}
\right].
\label{e:Omega_EdS}
\end{equation}
Choose $\ain \ll 1$ such that perturbations $\dlin(k,\ain)$ are
linear.  Since $D(a) = a$, $\dlin(k,a) = \dlin(k,\ain)a/\ain$.  SPT
expands the solution to the nonlinear evolution equations in powers of
the linear density contrast, $\delta(\vec k, a) = \sum_{n=1} a^n
\delta_n(\vec k)$, where $\delta_n(\vec k)$ is a mode-coupling
integral over the product of $n$ $\dlin$s, $\delta_n(\vec k) = \int
d^3p_0 \ldots d^3p_{n-1} \dirac(\vec k - \sum \vec p_i) F_n(\vec p_0,
\ldots, \vec p_{n-1})$ $\times \dlin(\vec p_0,\ain) \ldots \dlin(\vec
p_{n-1},\ain)$, and the $F_n$ are determined by
Eqns.~(\ref{e:continuity}-\ref{e:euler}) as in Ref.~\cite{Carlson09}.
The matter power spectrum $P(k,a)$ is then
\begin{eqnarray}
&(2\pi)^3& \dirac(\vec k + \vec k') P(k,a)
=
\left< \delta(\vec k) \delta(\vec k') \right>
\label{e:P_SPT}
\nonumber\\
&=& 
\!\!\!
\frac{a^2}{\ain^2} \left< \delta_1(\vec k) \delta_1(\vec k')\right>
+
2 \frac{a^4}{\ain^4} \left< \delta_1(\vec k) \delta_3(\vec k') \right>
\nonumber\\
&\quad&
+
\frac{a^4}{\ain^4} \left< \delta_2(\vec k) \delta_2 (\vec k') \right>
+
\ldots
\nonumber\\
&=&
\!\!\!
(2\pi)^3 \dirac(\vec k + \vec k')
\left[
  \Plin 
  + \Ponethree + \Ptwotwo 
  + \ldots 
\right]\quad
\end{eqnarray}
where $\Plin$ is the linear power spectrum, and the next-order
(``1-loop'') nonlinear corrections are given by~\cite{Makino92}:
\begin{eqnarray}
\Ponethree
&=&
\frac{k^3\Plin(k)}{1008\pi^2}
\int_0^\infty dr \Plin(kr)
\bigg[
  \frac{12}{r^2} - 158 + 100r^2 
\nonumber\\
&\quad&
- 42r^4 + \frac{3(r^2-1)^3(7r^2+2)}{r^2} \ln\left|\frac{1+r}{1-r}\right|
\bigg]
\label{e:Ponethree}
\\
\nonumber\\
\Ptwotwo
&=&
\frac{k^3}{392\pi^2}
\!\!
\int_0^\infty \!\!\! dr \Plin(kr)
\!\!
\int_{-1}^1 \!\!\! dx \Plin(k\sqrt{1 \! + \! r^2 \! - \! 2rx}) 
\nonumber\\
&\quad&
\times
\frac{(3r+7x-10rx^2)^2}{(1+r^2-2rx)^2}.
\label{e:Ptwotwo}
\end{eqnarray}
Here the dependence $\Plin(k,a) = \Plin(k,\ain) a^2/\ain^2$ upon $a$
has been suppressed.  The above approach can easily be extended to
higher-order terms (the ``2-loop'' terms) as well as to the velocity
power spectrum and the density-velocity cross power spectrum.

\subsection{Scale-Independent Growth}
\label{subsec:scale-independent_growth}

Now let us consider a universe which has a homogeneous component with
arbitrary equation of state in addition to CDM and baryonic matter,
resulting in a scale-independent growth factor $D(a)$.  This
homogeneous component may include a non-clustering dark energy as well
as a radiation component whose energy density is small enough that its
clustering may be neglected.  If we make the replacement $\varphi_1
\rightarrow \varphi_1 / f$ with $f = d\log D / d \log a$, and we
change the time variable in Eq.~(\ref{e:eom_compact}) from $\log a$ to
$\log[D(a)/D(\ain)]$, then the evolution matrix becomes
\begin{equation}
{\mathbf \Omega}
=
\left[
  \begin{array}{cc}
    1            & -1  \\
    -\frac{3 \Omegam(a)}{2f^2} & \frac{3 \Omegam(a)}{2f^2}
  \end{array}
\right].
\label{e:Omega_wCDM}
\end{equation}
If the equation of state parameter does not differ too much from $-1$,
then to reasonable precision, $f(a) \approx
\Omegam(a)^{0.55}$~\cite{Linder05}.  It follows that
$\Omegam(a)/f(a)^2 \approx \Omegam^{-1/10}$, which can itself be
approximated as unity at the $\approx 10\%$ level for $\Omegam(a) \geq
0.3$, and the evolution matrix can be approximated by its EdS
counterpart~(\ref{e:Omega_EdS}).  As a result, the $\Ponethree$ and
$\Ptwotwo$ corrections are given by
Eqs.~(\ref{e:Ponethree},~\ref{e:Ptwotwo}), the only difference being
the dependence of $\Plin(k,a) = \Plin(k,\ain) D(a)^2/D(\ain)^2$ upon
$a$.

\subsection{Massive Neutrinos and Time-RG}

Massive neutrinos cluster like cold matter on large scales but
free-stream out of bound structures on small scales.  Moreover, since
their velocities redshift away as the universe expands, the
free-streaming length scale changes with time. Thus the growth factor
depends on the wave number $k$ as well as $a$, and the method used in
Sec.~\ref{subsec:scale-independent_growth} to determine the $1$-loop
terms from their EdS counterparts breaks down.  (Rapidly evolving dark
energy models~\cite{Mortonson10} as well as scale-dependent fifth
forces from modified gravity are also inconsistent with the method of
Sec.~\ref{subsec:scale-independent_growth}.)  Scale-dependent growth
introduces a $k$-dependence into $\Omega_{10}$, the source term of the
Poisson equation, since CDM and neutrinos cluster differently on
different scales.

Time-Renormalization Group perturbation theory generalizes
Eq.~(\ref{e:eom_compact}) to $\vec k$-dependent $\mathbf{\Omega}$ and
integrates directly to find the power
spectrum~\cite{Pietroni08}. Since Eqs.~(\ref{e:continuity},
~\ref{e:euler}) relate the time-derivative of the first-order
perturbations $\delta$ and $\theta$ to second-order terms, there is an
infinite tower of evolution equations for the power spectra:
\begin{eqnarray}
\frac{\partial \corrtwo{a}{b}}{\partial \log a} 
&=&
-\Omega_{ac} \corrtwo{c}{b}
-\Omega_{bc} \corrtwo{a}{c}
\label{e:eom_2pt}
\\
&\quad&
+\frac{a}{\ain}\gamma_{acd}\left<\varphi_c \varphi_d \varphi_b\right>
+\frac{a}{\ain}\gamma_{bcd}\left<\varphi_a \varphi_c \varphi_d\right>
\nonumber\\
\frac{\partial \left<\varphi_a\varphi_b\varphi_c\right>}{\partial \log a}
&=&
-\Omega_{ad}\corrthree{d}{b}{c}
-\Omega_{bd} \corrthree{a}{d}{c}
\label{e:eom_3pt}
\\
&\quad&
-\Omega_{cd} \corrthree{a}{b}{d}
+\frac{a \gamma_{ade}}{\ain} \! \corrfour{d}{e}{b}{c}
\nonumber\\
&\quad&
+\frac{a \gamma_{bde}}{\ain} \! \corrfour{a}{d}{e}{c}
+\frac{a \gamma_{cde}}{\ain} \! \corrfour{a}{b}{d}{e}
\nonumber
\end{eqnarray}
and so on, each equation relating the evolution of the $n$-point
correlation functions to the $(n+1)$-point correlation functions.  In
this formalism, linear theory corresponds to setting to zero the
bispectrum $\dirac(\vec k + \vec p + \vec q) B_{abc}(\vec k,\vec
p,\vec q,a)$, truncating this tower after Eq.~(\ref{e:eom_2pt}).
Time-RG uses the next level of approximation, allowing nonzero
$B_{abc}$ but setting to zero the trispectrum, the connected part of
the four-point correlation function.  Direct integration of the $\vec
k$-dependent evolution equations~(\ref{e:eom_2pt},~\ref{e:eom_3pt})
means that the assumptions of scale-independent growth and a
time-independent evolution matrix are no longer necessary.

Consider a universe with two matter fluids, a cold fluid representing
CDM and baryons as well as a warm fluid representing neutrinos.  From
now on let $\varphi_0$ and $\varphi_1$ in
Eqs.~(\ref{e:varphi_0},~\ref{e:varphi_1}) refer to density and
velocity divergence perturbations in the cold fluid alone, denoted by
the subscript ${\mathrm{cb}}$.  Since neutrinos do not cluster on
small scales, they are well-described by the linearized evolution
equations.  Their density contrast $\delta_{\nu\mathrm{,lin}}(k,a)$
can be found using a linear Boltzmann code such as
\camb~\cite{Lewis:1999bs}, based on \cmbfast~\cite{Seljak96,
  Zaldarriaga98, Zaldarriaga00}. Then $\Omega_{10}$, the source term
for the CDM and baryon velocity divergence, is given by
\begin{equation}
\Omega_{10}(k,a)
=
-\frac{3}{2} \Omegam(a) 
\left[
  \fcb 
  + 
  \fnu \frac{\delta_{\nu\mathrm{,lin}}(k,a)}{\delta_\mathrm{cb,lin}(k,a)}
\right],
\label{e:Omega10_nu}
\end{equation}
where $\fcb = \Omega_\mathrm{cb}/\Omegam$ and $\fnu = \Omega_\nu /
\Omegam$ are evaluated today.  Note that $\Omega_{10}$ uses the linear
$\mathrm{cb}$ density contrast rather than the nonlinear one;
Ref.~\cite{Lesgourgues09} shows that this approximation introduces an
error of only $\approx 0.1\%$.

Time-RG perturbation theory directly integrates the evolution
equations~(\ref{e:eom_2pt},~\ref{e:eom_3pt}) with the evolution matrix
given by Eqs.~(\ref{e:Omega00},~\ref{e:Omega11},~\ref{e:Omega10_nu})
and vertices given by Eqs.~(\ref{e:gamma010},~\ref{e:gamma111}).
Initial conditions are given by $\corrtwo{a}{b} = \Plin f^{a+b}$ and
$\corrthree{a}{b}{c} = 0$ evaluated at $\ain$ sufficiently small that
the perturbations are linear.  Calculations presented here use a
modified version of the \copter~code~\cite{Carlson09,COPTER} in which
\emph{(i)} the homogeneous evolution includes massive neutrinos and
dynamical dark energy, and \emph{(ii)} the linear perturbations are
interpolated from \camb~outputs.

\begin{figure*}[tb]
\begin{center}
\includegraphics[angle=270,width=3.3in]{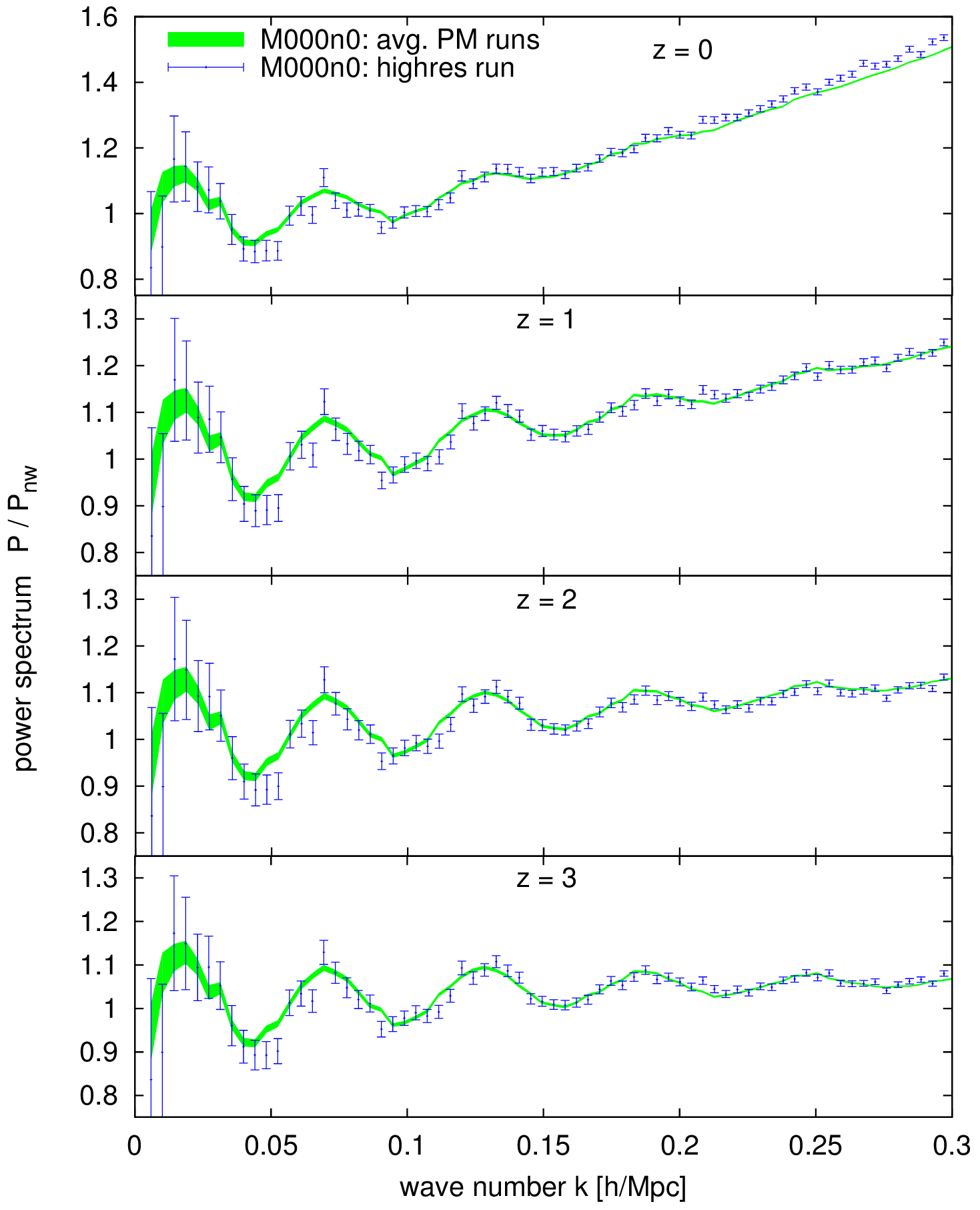}
\includegraphics[angle=270,width=3.3in]{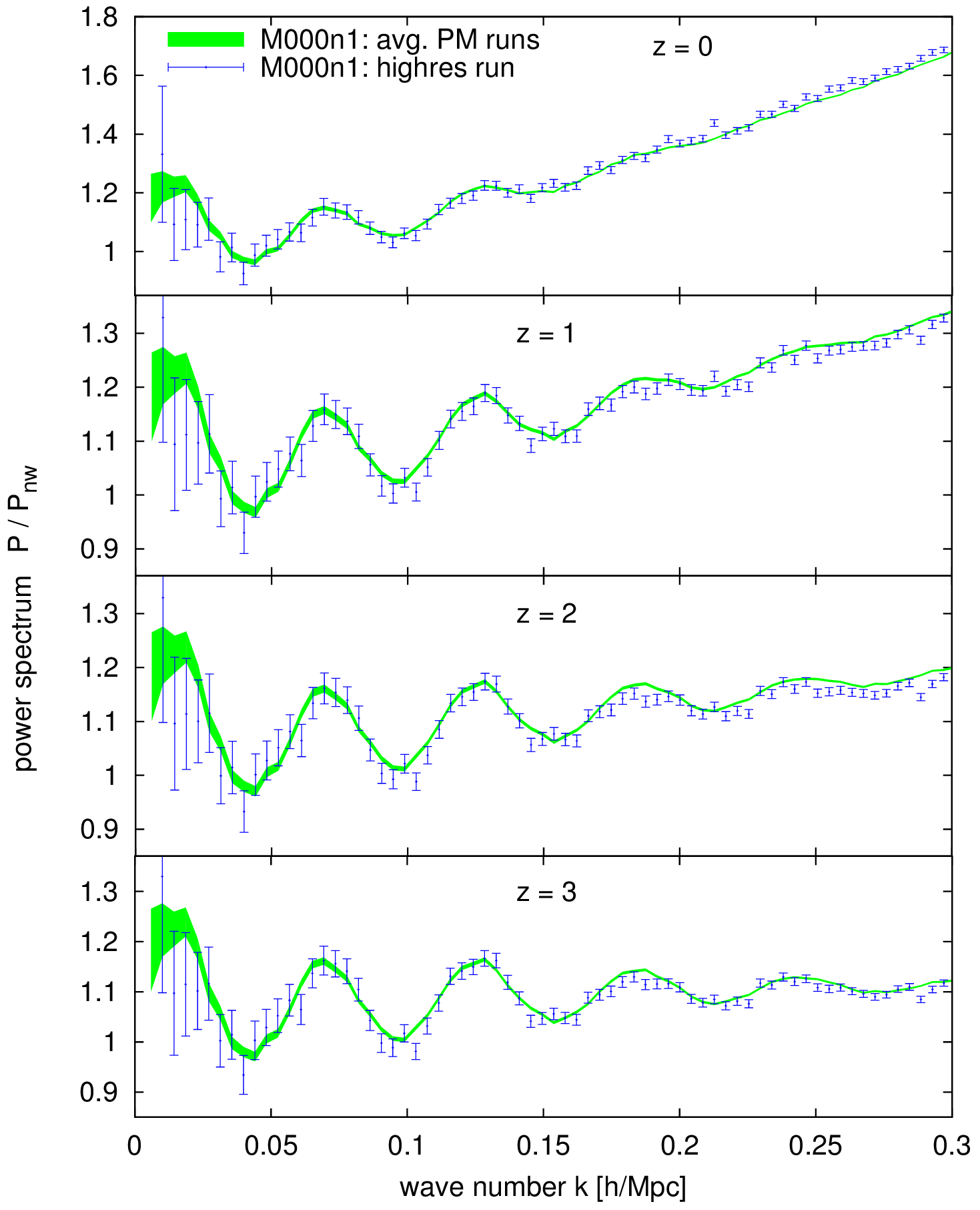}
\caption{Comparison between the average of $16$ PM realizations
  (green band) and a high resolution simulation (blue points). The agreement is
  very good up to $k \approx 0.3$~$h\mpcinv$ over the redshift range
  of interest. The left panel shows model M000n0 ($\Lambda$CDM) and
  the right panel, M000n1 ($\Lambda$CDM with massive neutrinos) of
  Table~\ref{tab:basic}, both divided by the no-wiggle power spectrum of
  Eq.~(\ref{e:Pnw_r_r}). 
  \label{f:highres}}
\end{center}
\end{figure*}

\section{N-body simulations}
\label{sec:n-body_simulations}

\subsection{Simulations with HACC}
\label{subsec:simulations_with_hacc}

In order to test the validity of the perturbation theory approach we
run a set of N-body simulations with the HACC
framework~\cite{Habib09,Pope10,Habib12}. HACC is a flexible N-body
code designed to exploit the diverse landscape of current and future
supercomputing architectures. HACC's design is centered around the
idea of breaking up the problem into long-range and short-range force
evaluations, keeping a highly optimized FFT-based long-range solver
the same on all architectures, while optimizing the short-range solver
for a specific target architecture. For hardware-accelerated systems,
such as those with graphics processing units (GPUs), particle-particle
particle mesh (P$^3$M) solvers can be easily optimized, while TreePM
methods are better suited for non-accelerated systems. HACC has been
shown to scale to the largest machines currently available. The
results shown in this paper have been obtained on the Blue Gene
systems Intrepid (BG/P) and Mira (BG/Q) at Argonne National Laboratory
and on Titan, a GPU-accelerated system at Oak Ridge National
Laboratory.

\begin{table*}
\begin{center} 
\caption{Parameters for the models investigated in this paper, where
  we use $\omega_\nu=\Sigma m_\nu/94$eV. \label{tab:basic}} 
\begin{tabular}{ccccccccccccc}
\# &L [$h^{-1}$Mpc] & $\omega_{cdm}$ & $\omega_b$ & $\omega_{\nu}$ &
$n_s$ &  $\sigma_8$ & $h$ & $w_0$ & $w_a$ & $\Sigma m_\nu$[eV] &PM
&High-res \\ \hline  
M000n0 & 1491.0 & 0.1109 & 0.02258 & 0.0 & 0.9630 & 0.8000 & 0.7100
&-1.0 & 0.0 & 0.0 & 16 & 1\\ 
M000n1 & 1491.0 & 0.1009 &  0.02258 & 0.010 & 0.9630 & 0.8000 & 0.7100
&-1.0 & 0.0 & 0.94 &16 & 1\\ 
M000n2 & 1491.0 & 0.1099 & 0.02258 & 0.001 &0.9630 & 0.8000 & 0.7100
&-1.0 & 0.0 & 0.094 & 16 & 0\\ 
M001n0 & 1295.1 & 0.1246 & 0.02261 & 0.0 & 0.9611 & 0.8778 & 0.6167 &
-0.7 & 0.6722 & 0.0 & 16 & 1\\  
M001n1 & 1295.1 & 0.1216 & 0.02261& 0.003 & 0.9611 & 0.8778 & 0.6167 &
-0.7 & 0.6722 & 0.282 & 16 & 0\\ 
M002n1 & 1224.9 & 0.0981 & 0.02283 & 0.003 & 0.8722 & 0.7789 & 0.5833
& -1.167  & 1.15 & 0.282 & 16 & 0 
\end{tabular}
\end{center}
\end{table*}

We ran a suite of N-body simulations covering different cosmological
models as summarized in Table~\ref{tab:basic}. Our investigations in
Section~\ref{sec:testing_perturbation_theory} start with a
$\Lambda$CDM model (M000) to set a well-tested baseline for studying
the range of validity of the different perturbation theories. For this
model we analyze one high-resolution simulation, evolving 3200$^3$
particles in a (2100~Mpc)$^3$ volume with a force resolution of
6.6~kpc.  The starting redshift of the simulation is $z_{\rm in}=200$
and the Zel'dovich approximation~\cite{Zeldovich:1969sb} is used to
set up the initial conditions. In order to obtain good statistics on
large scales, we also carry out a set of sixteen particle-mesh (PM)
simulations, evolving 512$^3$ particles on a 1024$^3$ uniform grid.
We use the high-resolution simulation to check that the PM simulations
yield accurate results up to the scales we are testing the different
perturbation theory approaches. As detailed in Table~\ref{tab:basic}
we have high-resolution simulations for three of the models, the
$\Lambda$CDM model with and without massive neutrinos and one model
with a time varying dark energy equation of state (all three
simulations evolving 3200$^3$ particles in a (2100~Mpc)$^3$ volume).
For all models we generate sixteen PM runs with the same
specifications as given above. Figure~\ref{f:highres} shows the
comparison of the high-resolution simulations with and without
neutrinos with the average power spectrum from the PM simulations. The
agreement is very good out to $k\sim0.3~h\mpcinv$, the maximum value
for which we compare our results with higher order perturbation theory
results presented in the next section.

In order to carry out simulations beyond $w$CDM models, we implement
some new features into HACC, namely a time varying equation of state
parameterized by $(w_0,w_a)$ via Eq.~(\ref{e:cpl}), as well as the
addition of massive neutrinos. As described in more detail below, we
treat neutrinos in an approximate way -- the perturbative results can
be used to estimate how well the approximations work, at least on
large and quasi-nonlinear scales. In the following, we provide a brief
description of our neutrino and dynamical dark energy implementations
within HACC.

\subsection{Neutrino Treatment and Dynamical Dark Energy}

The impact of dynamical dark energy and neutrinos on the simulated
matter power spectrum are taken into account by \emph{(i)} modifying
the initializer and \emph{(ii)} including both effects in the
background evolution. We do not model the interactions of massive
neutrino fluctuations with the dissipationless matter fluctuations
during the simulations, i.e., we run HACC as a gravity-only code with
a single species representing the sum of CDM and baryons. The total
matter power spectrum is constructed from the nonlinear CDM+baryon
power spectrum and the (linear) massive neutrino power spectrum from
{\tt{CAMB}} at the redshift of interest:
\begin{equation}
\label{e:pk}
P(k,a)=
\left[ \fcb \sqrt{\Pcb(k,a)} + \fnu \sqrt{P_\nu(k,a)}\right]^2.
\end{equation}
This approach is reasonable since neutrinos do not cluster strongly on
small scales, and has been adopted in previous numerical work, see,
e.g., Ref.~\cite{Agarwal11}, as well as in perturbation
theory~\cite{Saito08}.

In order to set up initial conditions for the HACC simulations we first
determine the shape and normalization of the total power spectrum
at $a=1$ ($z=0$) using linear theory:
\begin{equation}
P_{\rm total}(k,a=1) = {\rm A} k^{n_s} T_{\rm total}^2(k,a=1), 
\label{e:pktot}
\end{equation}
with $n_s$ being the primordial spectral index,
\begin{eqnarray}
&&T_{\rm total}(k,a=1) = \nonumber\\
&&\qquad\qquad f_{\rm cb} T_{\rm cb}(k,a=1)  + f_{\nu}
T_{\nu} (k,a=1),\\ 
&&T_{\rm cb}(k,a=1) = \nonumber\\
&&\qquad\qquad f_{\rm b} T_{\rm b}(k,a=1) + f_{\rm CDM}T_{\rm CDM}
(k,a=1), \quad 
\end{eqnarray}
and using an associated $\sigma_8$ normalization at $z=0$ which
implicitly defines the value of the amplitude coefficient, ${\rm A}$.
A scale-independent CDM-like growth function, $D(a)$, is then used to
move the $P_{\rm cb}$ piece of the power spectrum back to the initial
redshift, $z_i$, and to set the initial particle positions and
velocities for our single species code representing both CDM and
baryons.  This growth function takes into account all species in the
homogeneous background which makes comparison to high redshift linear
theory outputs from {\tt{CAMB}} more direct; radiation-like terms are
kept because they change the amplitude by several percent at $z \sim
100$. It does not however, take into account the scale-dependence that
neutrinos and baryons would contribute, therefore our terminology
``CDM-like''.

Our homogeneous background definitions assume that massive neutrinos,
if present, are massive enough to be matter-like at $z=0$. The
following equations are written for both massless and massive
neutrinos, though only one or the other may be present, not both:
\begin{eqnarray}
\Omega_r & = & \frac {2.471 \times 10^{-5}} {h^2} \left( \frac {T_{\rm
      CMB}} {2.725^{\circ} {\rm K}} \right)^4, \\ 
f_{\nu,r}^{\rm massless} & = &
\frac{7}{8}\left(\frac{4}{11}\right)^{4/3} N_{\rm eff}^{\rm massless},
\\ 
f_{\nu,r}^{\rm massive} & = &
\frac{7}{8}\left(\frac{4}{11}\right)^{4/3} N_{\rm eff}^{\rm massive},
\\ 
\Omega_{\nu}(a) & = & {\rm max}(\Omega_{\nu}a^{-3}, f_{\nu,r}^{\rm
  massive}\Omega_ra^{-4}), \\ 
H^2(a)/H^2_0 & = & \Omega_{cb}a^{-3} + (1+f_{\nu,r}^{\rm
  massless})\Omega_ra^{-4} + \Omega_{\nu}(a) \nonumber \\ 
& & + [1-\Omega_m-(1+f_{\nu,r}^{\rm massless})\Omega_r]\nonumber\\ 
&&\times a^{-3(1+w_0+w_a)}\exp[-3 w_a(1-a)]. \label{eq:hubble}
\end{eqnarray}

The scale independent CDM-like growth function is the linear
equivalent of the (nonlinear) gravity-only operator in HACC that uses
the equivalent definitions of the homogenous background as in
Eq.~(\ref{eq:hubble}).

The approximations used to incorporate the effects of baryons and
massive neutrinos are similar in that each has a scale-dependent
growth and would impart scale-dependence in the CDM growth.  The
shapes and amplitudes are also defined in terms of the linear power
spectra, and the goal is to produce accurate power spectra at low
redshifts with nonlinear clustering effects in the CDM and
baryons. The approximations for baryons and massive neutrinos differ
in that the mass of baryons is deposited on the dissipationless
gravitationally-interacting particles advanced by HACC, but the
massive neutrinos are only accounted for with linear theory, as
described above.

We would like to stress that using the total power spectrum from
{\tt{CAMB}} directly at the initial redshift to initialize the
particle positions and velocities would lead to inconsistent results
in a gravity-only N-body code: {\tt{CAMB}} accounts for baryon-photon 
coupling and the scale dependence in the growth function which
are absent in the N-body code.

In order to account for the presence of dynamical dark energy we
modify {\tt{CAMB}} as follows. For including dynamical dark energy, we
need to \emph{(i)} modify the equations describing the evolution of
the background which results in a change of the perturbations of dark
matter, radiation and neutrinos and \emph{(ii)} modify the equations
describing the density and velocity perturbations in the dark energy.
The modification to the background cosmology is trivially achieved by
modifying the equation describing the evolution of conformal time as a
function of the scale factor, using the evolution of dark energy
density [see Eq.~(\ref{eq:hubble})].  Modifying the equations
describing the perturbations of dark energy requires an expression for
the speed of sound. Consistent with a simple scalar field model, we
assume that this is the speed of light. As a result, perturbations of
dark energy develop only on the Hubble scale. A second issue is that
the equations describing the evolution of the velocity perturbations
include terms of the form $c_s^2/(1+w(a))$, where $c_s$ is the speed
of sound in the rest frame of dark energy. For those values of $(w_0,
w_a)$ for which the equation of state passes through $-1$, this
results in a singularity.  Assuming that at the crossing, this term is
small enough that the microscopic properties of dark energy do not
modify the power spectra, we replace the term in a small range around
the crossing by linearly interpolating between the values at the end
of the range, where this term is finite.

\section{Results and Discussion}
\label{sec:testing_perturbation_theory}

\subsection{The Regime of Validity of Perturbation Theory}

In general, the convergence properties of perturbation theories for
the evolution of matter fluctuations cannot be rigorously calculated.
Furthermore, at sufficiently nonlinear scales, the fluid approximation
to the Vlasov-Poisson equation itself breaks
down~\cite{Pueblas_etal_2009, Valageas_2011, Anselmi12,
  Pietroni_etal_2012}.  In the absence of a reliable internal test,
the best determination of the accuracy of higher-order perturbative
methods is direct comparison against N-body simulations.  Since the
scales of interest here are large enough that baryonic effects on the
growth are small, and sufficient force and mass resolution can be
easily attained, predictions for the power spectrum can be controlled
to accuracies of better than a percent by using N-body methods~(see,
e.g. Ref.~\cite{Heitmann10}).

\begin{figure*}[tb]
\begin{center}
\includegraphics[width=2.3in]{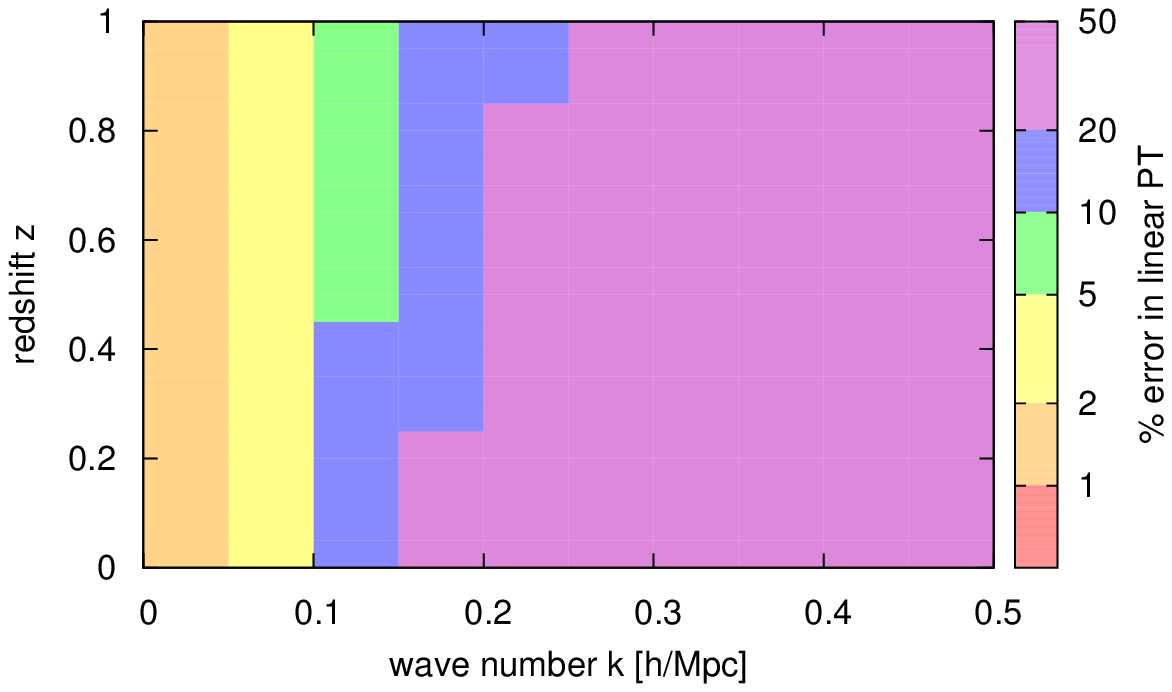}
\hskip0.0in
\includegraphics[width=2.3in]{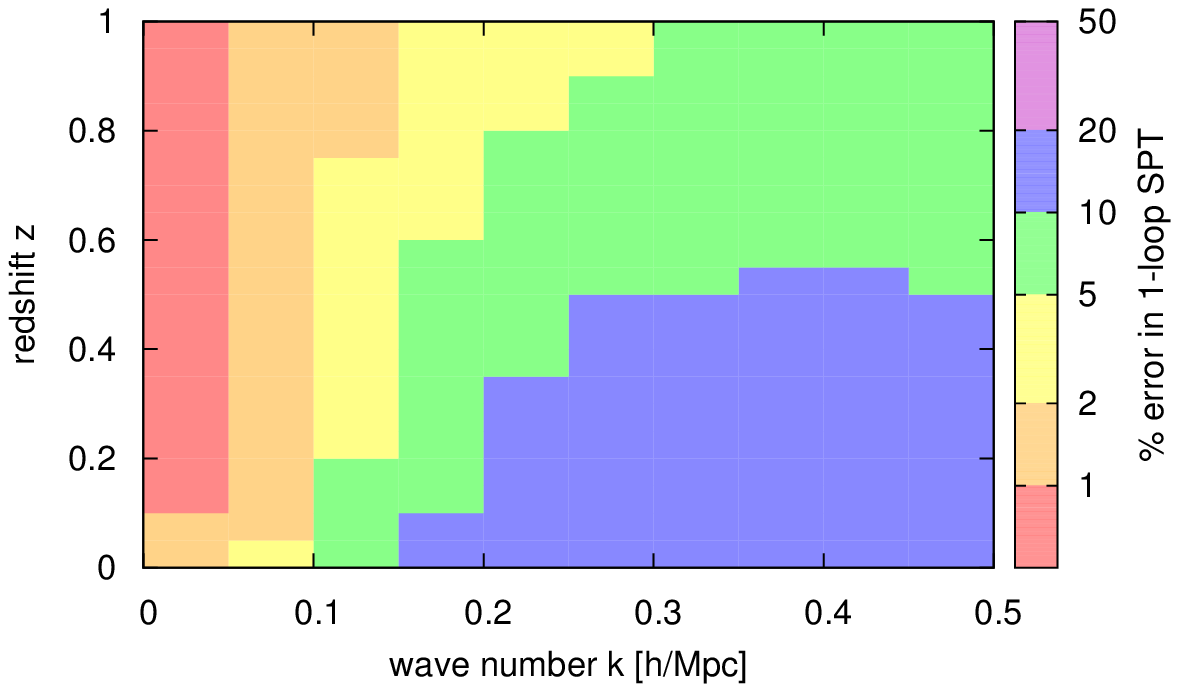}
\hskip0.0in
\includegraphics[width=2.3in]{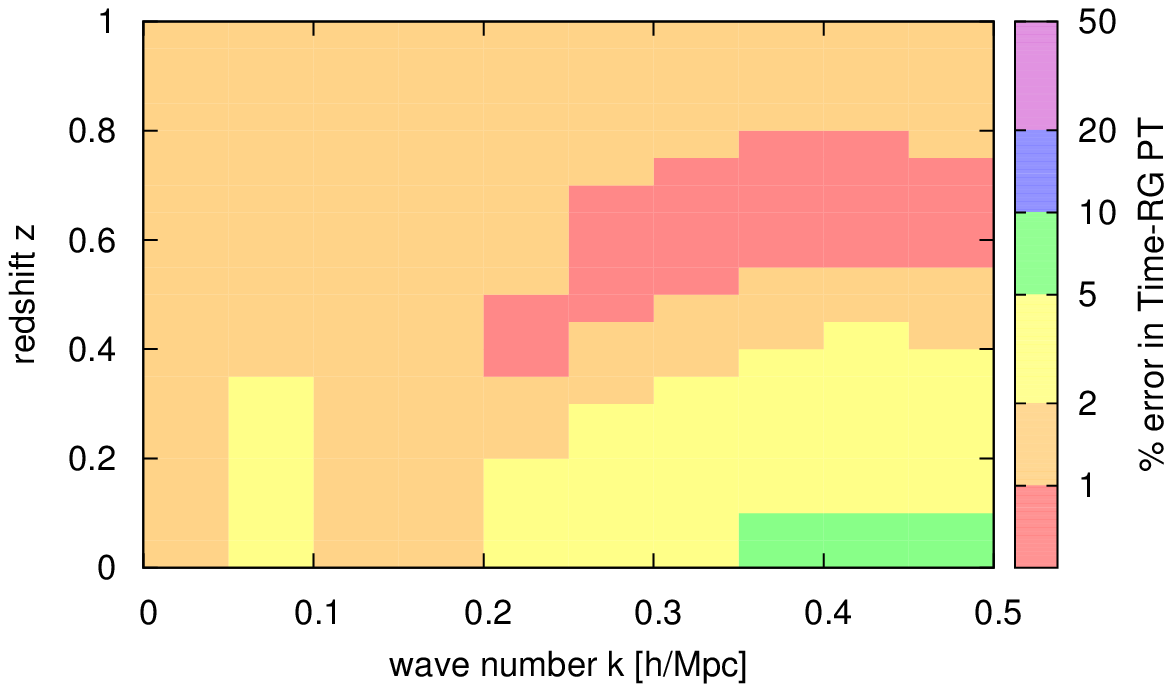}
\caption{Comparison of linear perturbation theory (left), 
  1-loop SPT (middle) and Time-RG (right) to the
  cosmic emulator of Ref.~\cite{Lawrence09} for dark energy with
  constant $w=-1.2$ and massless neutrinos.  
  The color of each bin corresponds to the maximum difference
  between perturbation theory and the emulator power spectrum.
  \label{f:comparison_const_w_r}}
\end{center}
\end{figure*}

We begin by testing our perturbative calculation for a cosmology with
constant dark energy equation of state and massless neutrinos.  For
this purpose, we use the cosmic emulator of Ref.~\cite{Lawrence09},
which uses Gaussian process modeling to interpolate the results of
$37$ high-resolution N-body simulations chosen to span the
cosmological model parameter space. (Previous comparisons of the
emulator and perturbation theory can be found in
Refs.~\cite{Anselmi12, Taruya12}.)
Figure~\ref{f:comparison_const_w_r} compares linear perturbation
theory, 1-loop SPT, and Time-RG to the power spectrum emulator, which
is accurate to $1\%$ for $z<1$ and $k < 1$~$h\mpcinv$.  (Full
2-loop SPT is sufficiently time-consuming that a similar calculation
would be difficult.)  As expected, 1-loop SPT performs significantly
better than linear perturbation theory, especially at high $z$.
Time-RG, which includes some 2-loop terms, is even more accurate.

\begin{figure*}[tb]
\begin{center}
\includegraphics[angle=270,width=3.3in]{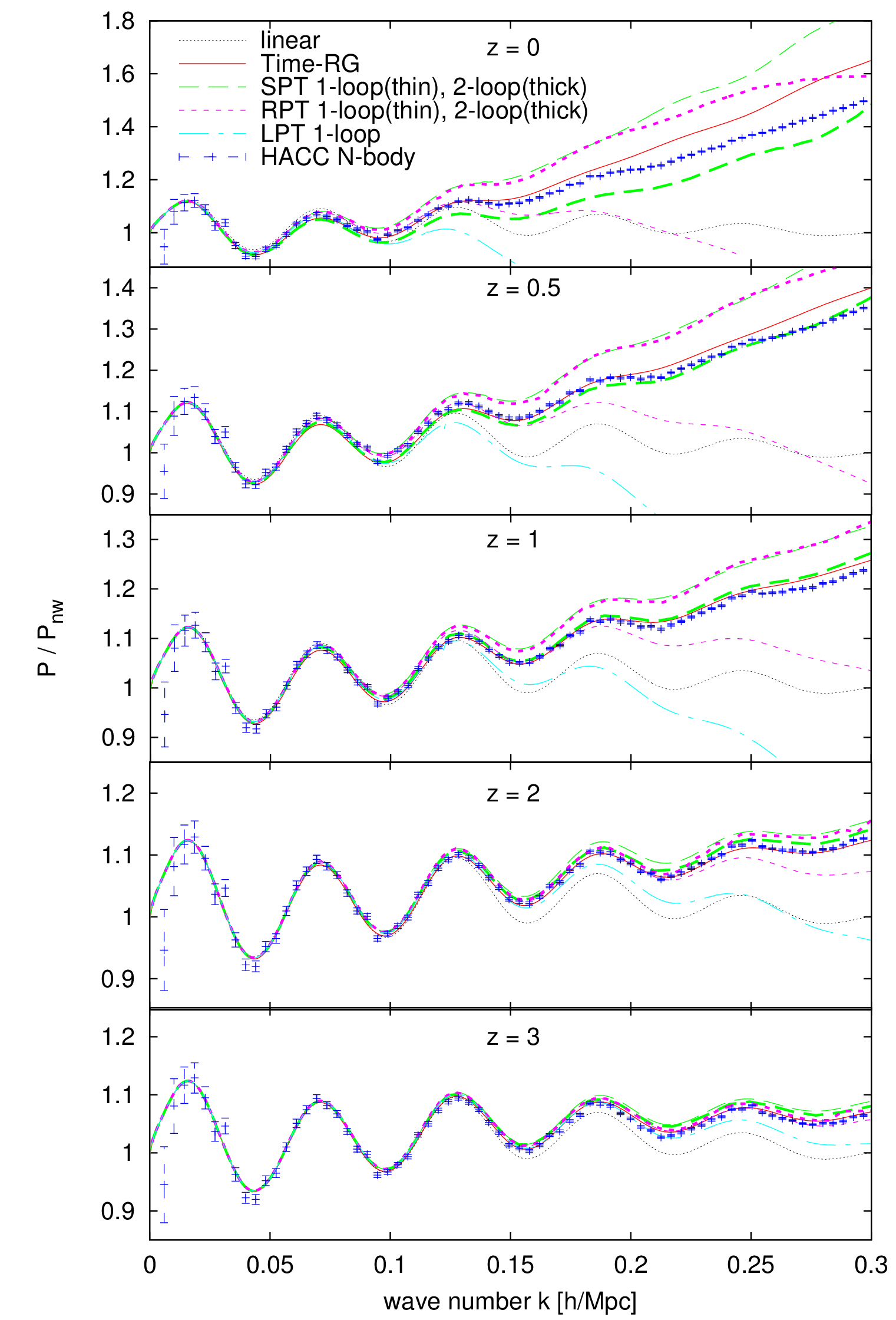}
\hskip0.2in
\includegraphics[angle=270,width=3.3in]{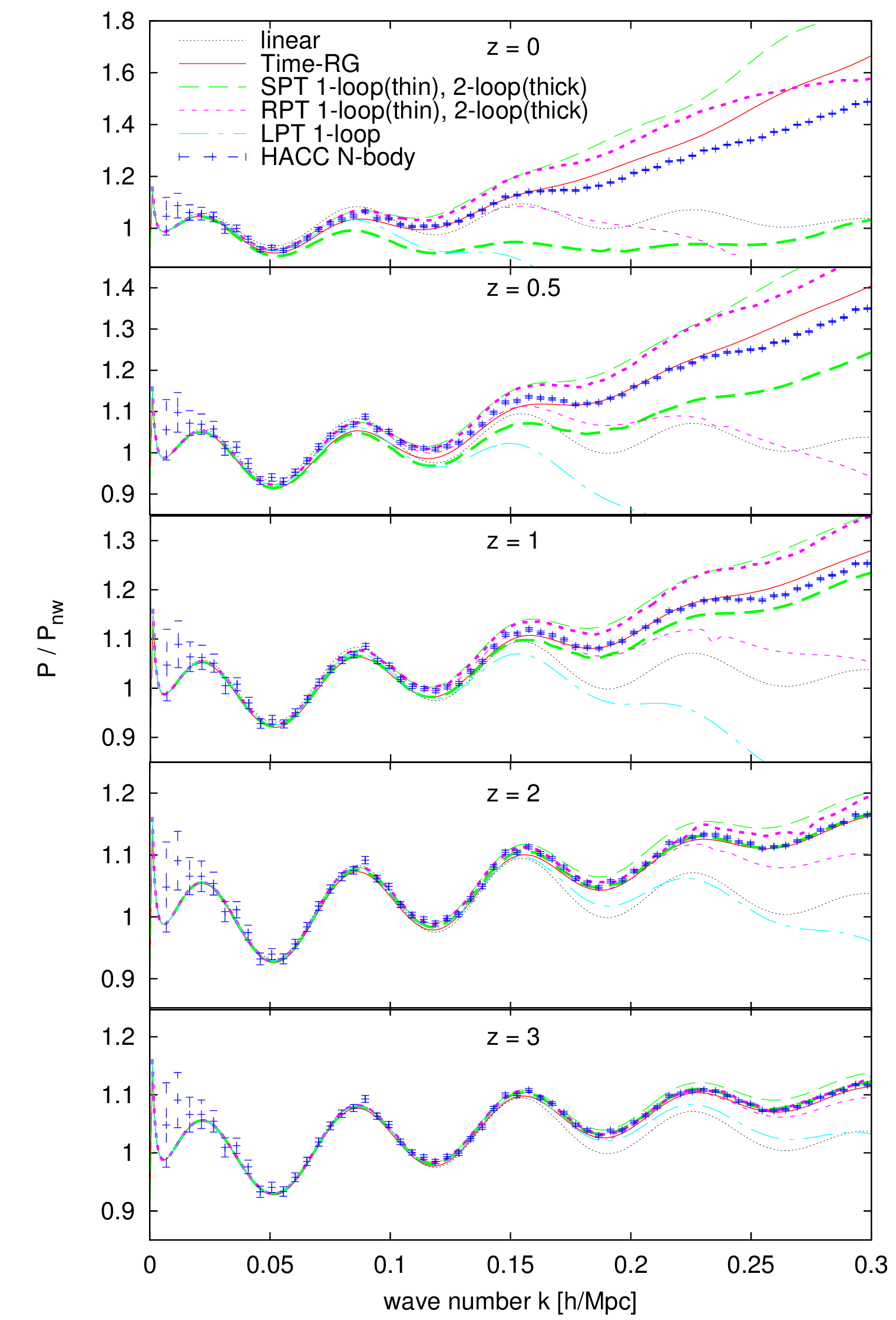}
\caption{Power spectrum ratios for linear theory and four different
  higher-order perturbation theories compared with the HACC N-body
  power spectrum at 5 redshifts for models with massless neutrinos.
  For SPT and RPT, thin and thick lines denote 1-loop and 2-loop
  calculations, respectively. 
  Left:~$\Lambda$CDM, model M000n0.  Right:~Early dark
  energy, model M001n0.  The wavenumbers at which the perturbation
  theory results deviate from the simulations by $1\%$ and $2\%$
  are shown in Table ~\ref{t:PTerror}.  \label{f:allPT_r}}
\end{center}
\end{figure*}

\begin{figure}
\begin{center}
\includegraphics[angle=270,width=3.3in]{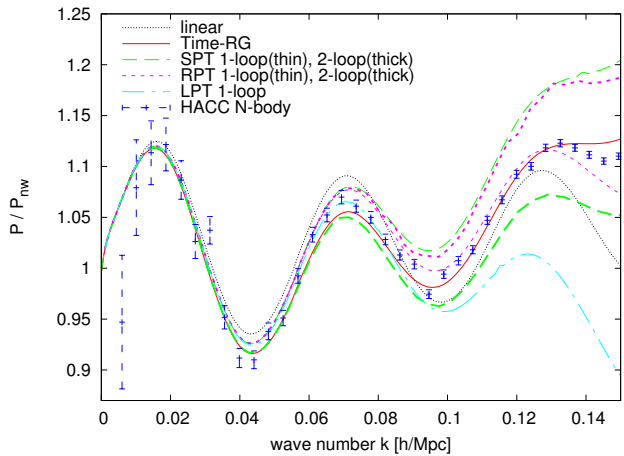}
\caption{
  Power spectra at $z=0$ for the M000n0 model from Fig.~\ref{f:allPT_r},
  showing nonlinear suppression of power at intermediate scales.
  \label{f:allPT_lowk}
}
\end{center}
\end{figure}

\begin{table}
\begin{center}
\caption{Wave number $k$ [h \mpcinvtxt] below which each perturbation theory
  is accurate to $1\%$ (or $2\%$) in models with massless neutrinos. 
  \label{t:PTerror}} 
\tabcolsep=0.05cm
\begin{tabular}{lcc|ccccccc}


\rotatebox{90}{\hskip-0.2in Model} & $z$  & Acc. & Linear       & \multicolumn{2}{c}{SPT}     &
\multicolumn{2}{c}{RPT}     & Time-RG      & LPT\\ 
      &      &      &              & 1-loop       & 2-loop       & 1-loop
      & 2-loop       &              &    \\ 
\hline
\rotatebox{90}{\hskip-0.5in M000n0} 
       & $0$ & $1\%$ &0.076 & 0.084 & 0.093 & 0.14 & 0.084 & 0.15 & 0.093 \\
       &     & \Tabg{$2\%$} & \Tabg{0.1} & \Tabg{0.11} & \Tabg{0.1} & \Tabg{0.15} & \Tabg{0.11} & \Tabg{0.19} & \Tabg{0.1} \\
       & $1$ & $1\%$ &0.13 & 0.12 & 0.24 & 0.2 & 0.13 & 0.28 & 0.14 \\
       &     & \Tabg{$2\%$} & \Tabg{0.14} & \Tabg{0.15} & \Tabg{0.3} & \Tabg{0.21} & \Tabg{0.17} & \Tabg{0.97} & \Tabg{0.15} \\

\hline
\rotatebox{90}{\hskip-0.5in M001n0} 
       & $0$ & $1\%$ &0.078 & 0.11 & 0.058 & 0.15 & 0.11 & 0.17 & 0.087 \\
       &     & \Tabg{$2\%$} & \Tabg{0.078} & \Tabg{0.11} & \Tabg{0.068} & \Tabg{0.15} & \Tabg{0.13} & \Tabg{0.18} & \Tabg{0.097} \\
       & $1$ & $1\%$ &0.12 & 0.14 & 0.16 & 0.19 & 0.16 & 0.26 & 0.14 \\
       &     & \Tabg{$2\%$} & \Tabg{0.16} & \Tabg{0.17} & \Tabg{0.29} & \Tabg{0.2} & \Tabg{0.19} & \Tabg{0.32} & \Tabg{0.15} \\

\end{tabular}
\end{center}
\end{table}

In order to improve the accuracy of the power spectrum computation and
to test perturbation theory for models with $w_a \neq 0$, we run
high-resolution N-body simulations, shown in Table~\ref{tab:basic}. Model
M000n0 is a standard $\Lambda$CDM model, while M001n0 is an early dark
energy model in which $w(z)$ evolves rapidly, allowing the dark energy
to be a substantial fraction of the total energy density at the time
of recombination.  Figure~\ref{f:allPT_r} compares linear theory,
Time-RG, Standard Perturbation Theory (SPT), Renormalized Perturbation
Theory (RPT), and Lagrangian Resummation Perturbation Theory (LPT) to
the HACC N-body power spectrum for $\Lambda$CDM and early dark
energy. RPT is a resummed alternative to the SPT discussed earlier.
LPT is formulated in terms of particle displacements, making it
particularly useful for observations in redshift space.

In order to present the results more clearly, we have divided the
power spectra by a smoothed ``no-wiggle'' power spectrum $\Pnw(k,z)$:
\begin{eqnarray}
\frac{1}{\Pnw(k,0)^{\nnw}}
&=&
\frac{1}{\left[\left(\frac{k}{k_0}\right)^{n_s-\enw} P_0 \right]^{\nnw}}
+
\frac{1}{\left[\left(\frac{k}{k_1}\right)^{n_s-3} P_1 \right]^{\nnw}},
\qquad
\label{e:Pnw_r_r}
\\
\Pnw(k,z) 
&=& 
D(z)^2 \Pnw(k,0).
\end{eqnarray}
Here $k_0 = 10^{-4}~h\mpcinv$, $k_1 = 0.3~h\mpcinv$, $\nnw = 0.4$, and
we have chosen $P_0$ and $P_1$ such that $\Pnw(k_0)/\Plin(k_0) \approx
\Pnw(k_1)/\Plin(k_0) \approx 1$.  We choose $\enw = 0$ except for
early dark energy models (we used $\enw =0.2$ for models M001n0 and
M001n1; $\enw = 0.04$ for M002n1), in which dark energy is a
significant fraction $\sim 10\%$ of the total energy density even at
$z \sim 1000$, and the universe is never completely matter-dominated.
$D(z)$ is the scale-independent growth factor found by setting $\delta_\nu=0$
in Eq.~(\ref{e:Omega10_nu}).

Figure~\ref{f:allPT_r}~(left) extends the result of Ref.~\cite{Carlson09}
to higher redshifts, while Fig.~\ref{f:allPT_r}~(right) is a new
result.  Table~\ref{t:PTerror} summarizes the results, showing the $k$
at which each perturbation theory begins to differ from simulations by
$1\%$ or $2\%$.  Since the simulated power spectra are noisy at low
$k$, in practice we added the $1\%$ or $2\%$ errors to the $3\sigma$
statistical uncertainty of the N-body power spectra.

From Fig.~\ref{f:allPT_lowk} we see that higher-order perturbation theories
correctly predict the power spectrum falling below linear theory in
the range $0.05~h\mpcinv \lesssim k \lesssim
0.1~h\mpcinv$ as power moves from large scales to small scales.
RPT and $1$-loop SPT predict the smallest dip, and  
Time-RG and $2$-loop SPT predict the largest, while the
N-body simulations prefer an intermediate value.  In the range
$0.1~h\mpcinv \lesssim k \lesssim 0.15~h\mpcinv$, Time-RG and $1$-loop RPT
most closely approximate the simulations, while SPT and LPT begin to
diverge from the other power spectra.  At the $2\%$ level, all of the
higher-order perturbative methods are accurate up to $k \approx
0.1~h\mpcinv$ at $z=0$.  

\begin{figure*}[tb]
\begin{center}
\includegraphics[angle=270,width=3.3in]{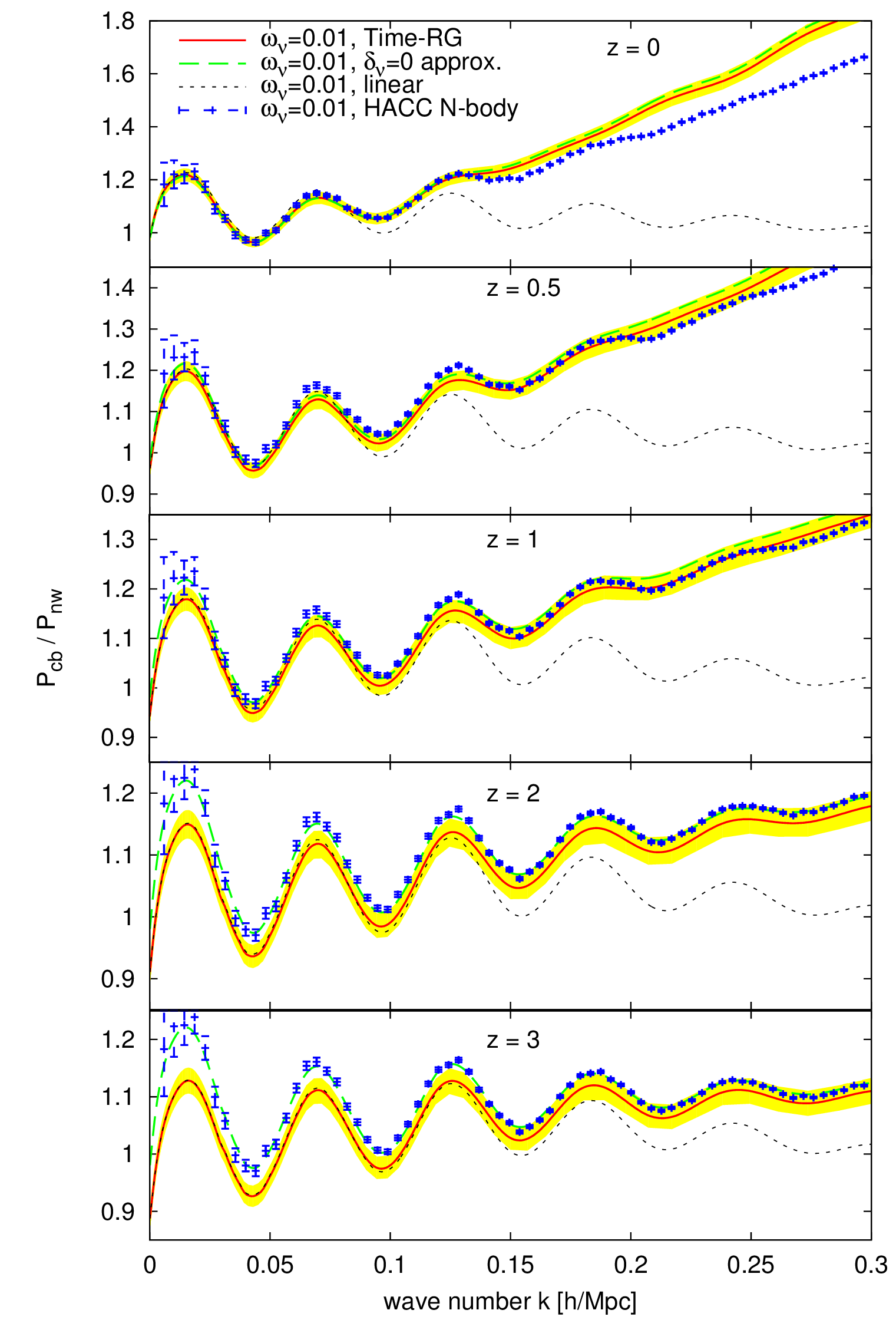}
\hskip0.2in
\includegraphics[angle=270,width=3.3in]{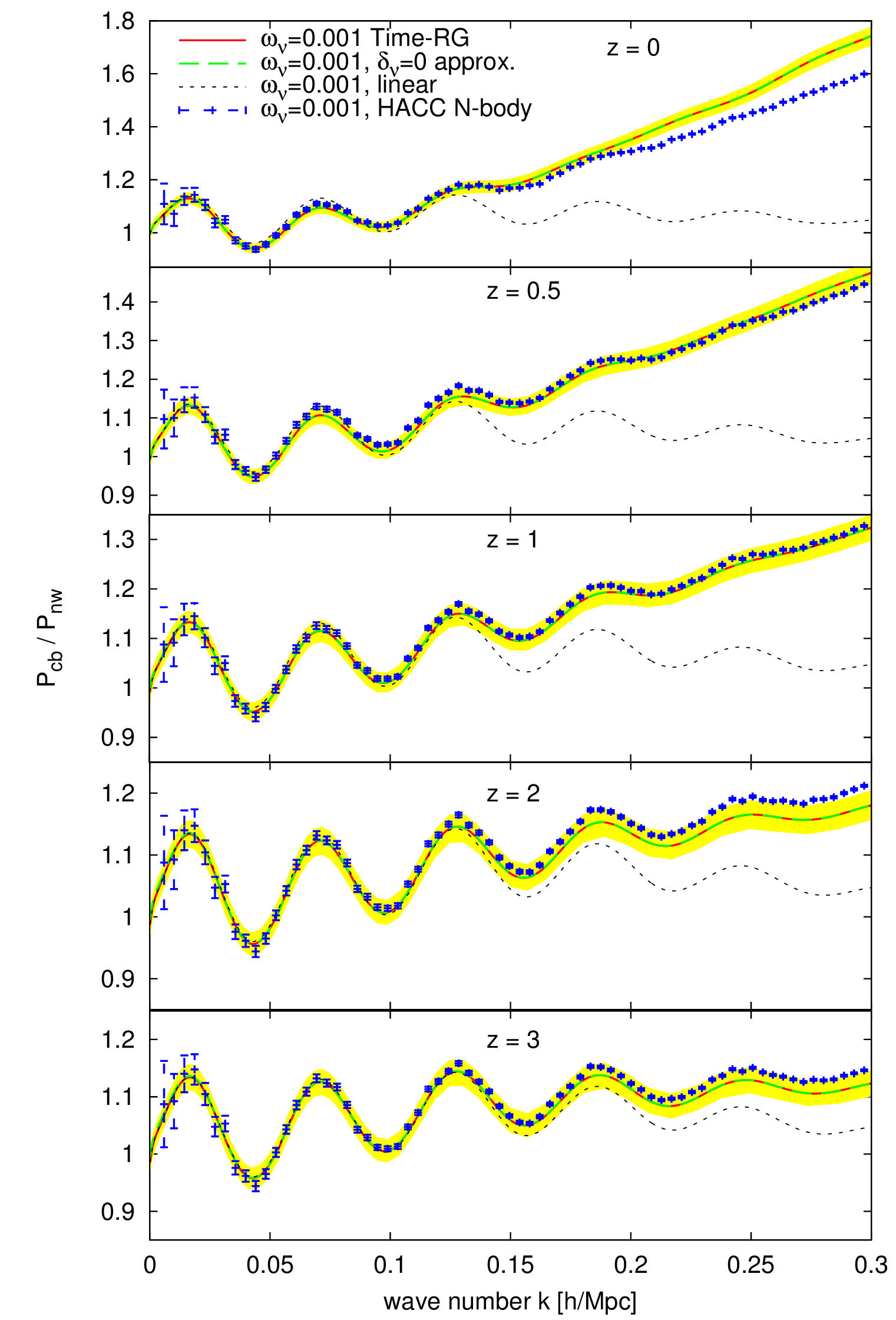}
\caption{Power spectra $P_{\rm cb}$ for models M000n1 (left) and M000n2
  (right), with a cosmological constant and massive neutrinos.  $\Pnw$
  is the no-wiggle power spectrum of Eq.~(\ref{e:Pnw_r_r}) associated with
  the linear power spectrum at $z=0$ in each case.  Shaded yellow
  regions show power spectra within $2\%$ of the corresponding Time-RG
  curves.  Power spectra are also shown with the neutrino
  density contrast $\delta_\nu$ set to zero (green dashed curves), the
  approximation used in the HACC simulations. See the text for further
  discussion. \label{f:m000n}}
\end{center}
\end{figure*}

In the models above, the N-body simulations (and therefore the
emulator results based on them) include all gravity effects
consistently. When neutrinos are added, this is no longer the case; as
discussed earlier, the gravitational potential of clustering of
massive neutrinos is not included in the simulations, as this effect
is expected to be small. However, a finite neutrino mass leads to a
suppression of the power spectrum, and therefore pushes the onset of
nonlinear effects to higher wave numbers. Thus, for cosmological
models obtained by changing a fraction of the dark matter energy
density to the energy density of massive neutrinos (at the same
low-$k$ amplitude), perturbation theory must continue to be valid in
at least the regime obtained above. Since Time-RG consistently
includes the massive neutrinos, we can use it to test the extensions
of HACC for massive neutrinos.

\begin{figure*}
\begin{center}
\includegraphics[angle=270,width=3.3in]{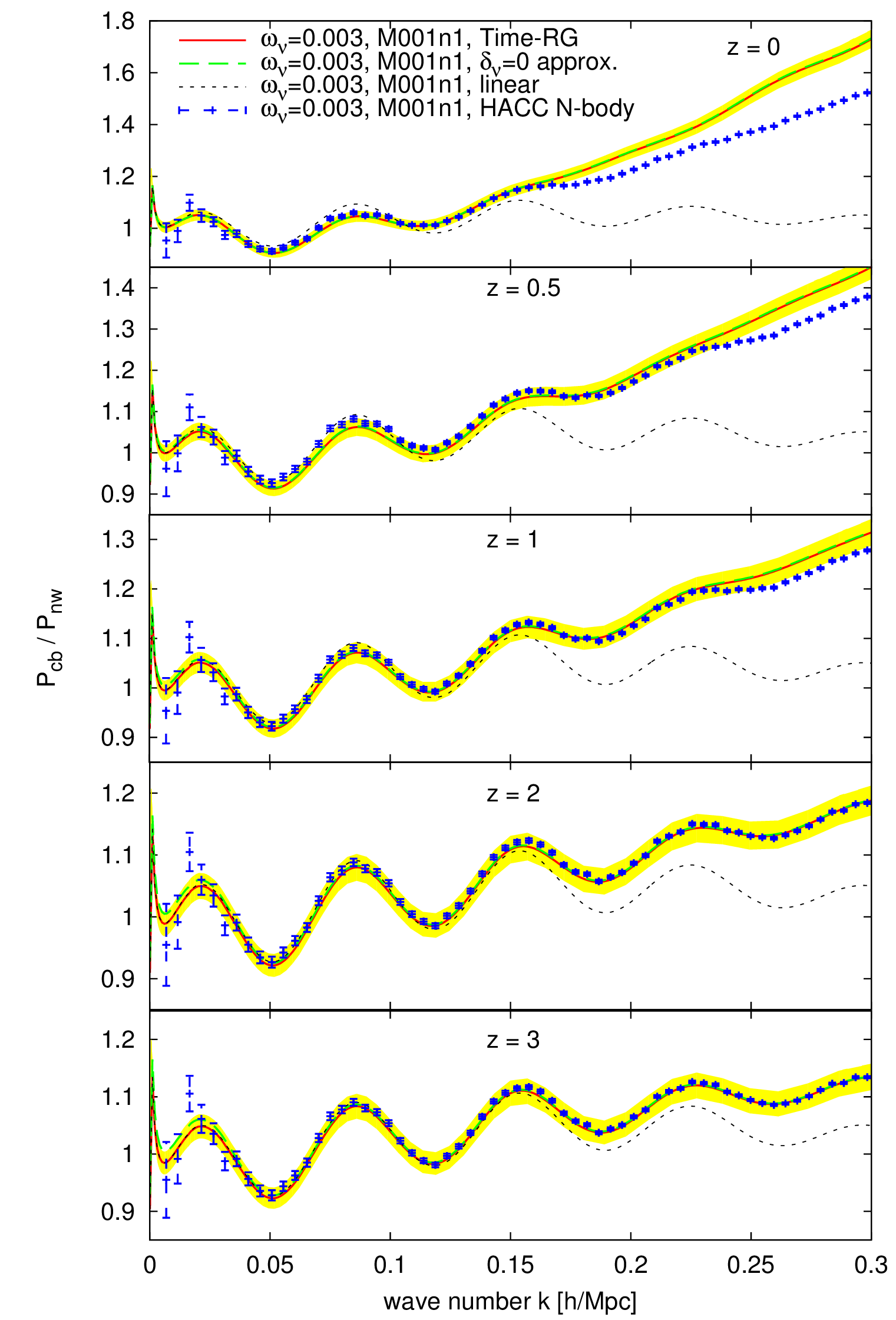}
\hskip0.2in
\includegraphics[angle=270,width=3.3in]{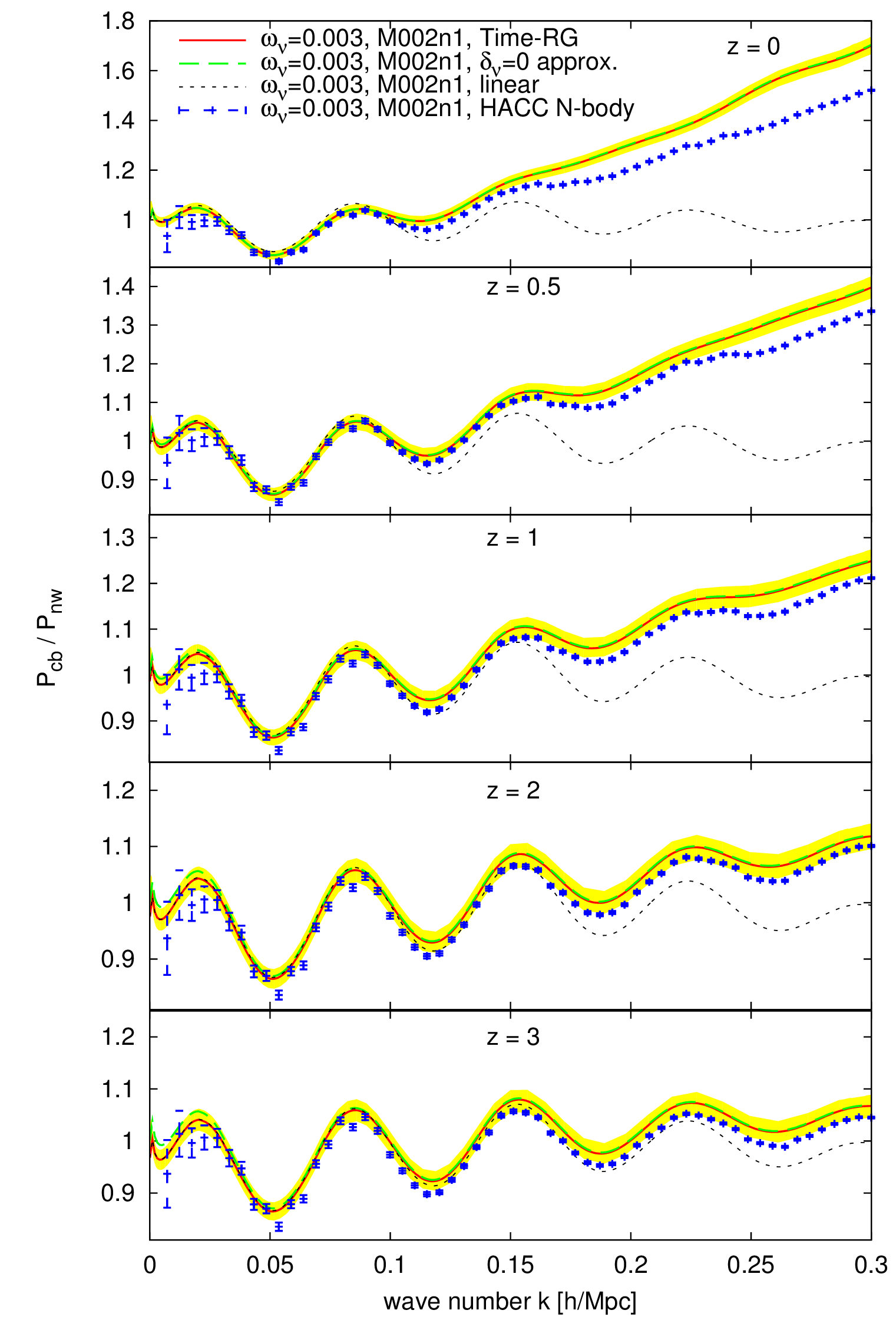}
\caption{Power spectra $P_{\rm cb}$ for the early dark energy models
  with massive neutrinos, M001n1 (left) and M002n1 (right), following
  the conventions of Fig.~\ref{f:m000n}. \label{f:nuEDE_r}}
\end{center}
\end{figure*}

\subsection{Massive Neutrinos: Perturbation Theory and HACC
  Simulations}  
We now consider cosmological models with massive neutrinos.  Since
Time-RG is the only perturbative method we consider that includes
neutrino masses, all perturbative calculations described below are
restricted to Time-RG.  Figure~\ref{f:m000n} shows the power spectra
of models M000n1 and M000n2, both of which have cosmological constants
and massive neutrinos.  The yellow shaded band around the Time-RG
curve identifies power spectra within $2\%$ of Time-RG.  When
$\omega_\nu$ is small as in Fig.~\ref{f:m000n}~(right), Time-RG agrees
with the HACC results to $2\%$ for $k \approx 0.2~h\mpcinv$ at $z=0$,
and for even higher $k$ at $z \geq 0.5$.  Since Fig.~\ref{f:m000n}~(right)
corresponds to $\Sigma m_\nu = 0.094$~eV, about half of the current upper 
bound, we expect Time-RG to be a good approximation in the most interesting
region of parameter space.

The results of Fig.~\ref{f:m000n}~(left) for model M000n1, in which
$\omega_\nu=0.01$ implies a significant neutrino fraction $\fnu =
7.5\%$ and mass $\Sigma m_\nu = 0.94$~eV, show a discrepancy between 
Time-RG and the approximate
treatment implemented in HACC even for $k<0.1~h\mpcinv$ for the
higher-$z$ cases considered.  Because the approximation does not
include neutrinos as a source for CDM+baryon growth, it is expected to
misestimate the power at large $\fnu$. 
Note, however, that the neutrino mass implied by this model is about
four times as high as the bound from~\cite{Ade_etal_2013xvi}.

At $z=0$, the N-body and
perturbative results are normalized correctly by construction,
agreeing well at the lowest $k$ values. 
However, at higher $z$, the Time-RG curve (red line) 
should be above the N-body curve (blue points) because the additional 
neutrino sourcing implies a larger growth
function, and hence smaller power spectrum at higher $z$ after
normalizing to $\sigma_8$ at $z=0$.
In order to display this effect and to estimate its
magnitude, we also carried out the Time-RG calculation with
$\delta_{\nu\mathrm{,lin}}$ set to zero in Eq.~(\ref{e:Omega10_nu}),
as shown in the long-dashed green curve in Fig.~\ref{f:m000n}~(left).
This $\delta_\nu=0$ curve is in very good agreement with the
approximate HACC power spectrum in the low-$k$ regime at all
redshifts, showing explicitly that the discrepancy between the Time-RG
and HACC results is due to the approximate treatment of neutrinos
discussed in Sec.~\ref{sec:n-body_simulations}. (Compensation for this
error at the linear level is possible, but we do not pursue it here.)
Moreover, a comparison between the Time-RG and $\delta_\nu=0$ curves
provides an estimate of the accuracy of this approximation, which is
better than $1\%$ for $z=0$ up to $k=0.16~h$\mpcinvtxt and better than
$1.5\%$ up to $k=0.27~h\mpcinv$ (although at this point, Time-RG is
clearly wrong). This applies to $\omega_\nu=0.01$, so the error will
be several times smaller for lower $\omega_\nu$.

Finally, Fig.~\ref{f:nuEDE_r} shows power spectra for two different
early dark energy models with massive neutrinos.  For both models, 
Time-RG works quite well, agreeing to
mostly better than $2\%$ with the N-body results up to
$k=0.17~h\mpcinv$ at $z=0$.

One possible application of our perturbative results is to combine
them with N-body simulations in order to obtain an accurate power
spectrum calculation over the greatest possible range of scales.
Higher-order perturbation theories such as Time-RG are accurate up to
$k = 0.05 - 0.1~h\mpcinv$, as we have confirmed over a large range of
dark energy equations of state and neutrino masses.  These large
scales are precisely where simulations can have some difficulties due
to their finite box sizes. Thus by combining Time-RG calculations with
those of HACC, it is possible to predict the power spectrum from
horizon scales to $k \gtrsim 1~h\mpcinv$.

\subsection{Impact on `Observables':~Exploring Physics Beyond wCDM using  
  Perturbation Theory} 
\label{sec:observables}
\subsubsection{Dynamical Dark Energy}

Growth of structure depends on the dark energy equation of state.  In
the linear regime the dominant effect will be a scale-independent
change to the growth factor $D(a)$. Since we normalize power spectra
using $\sigma_8$ at $z=0$, this effect will be most noticeable at
higher redshifts.  Meanwhile, nonlinearities may introduce a
scale-dependent change at larger $k$.

\begin{figure}[htb]
\begin{center}
\includegraphics[angle=270,width=3.3in]{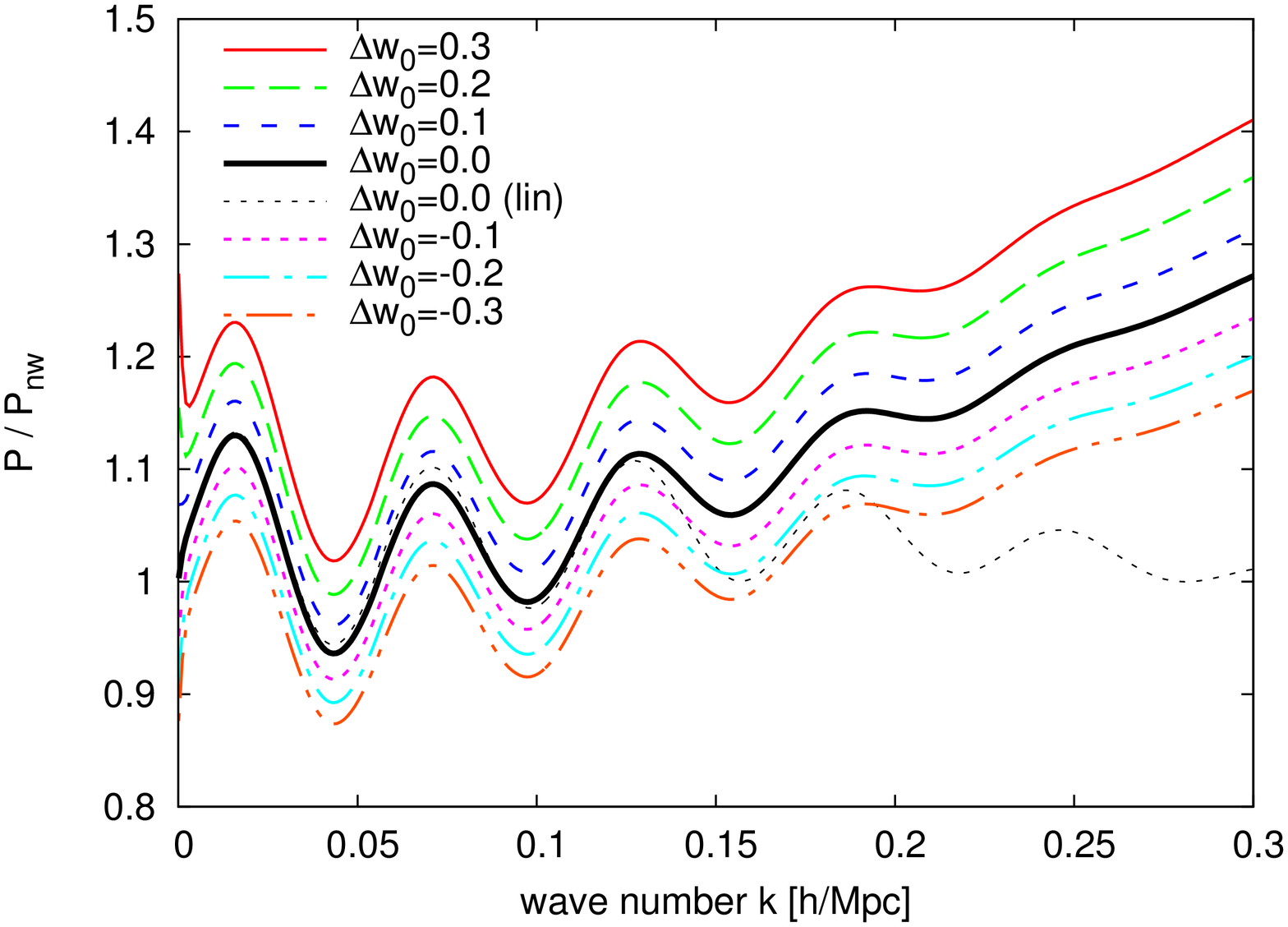}
\includegraphics[angle=270,width=3.3in]{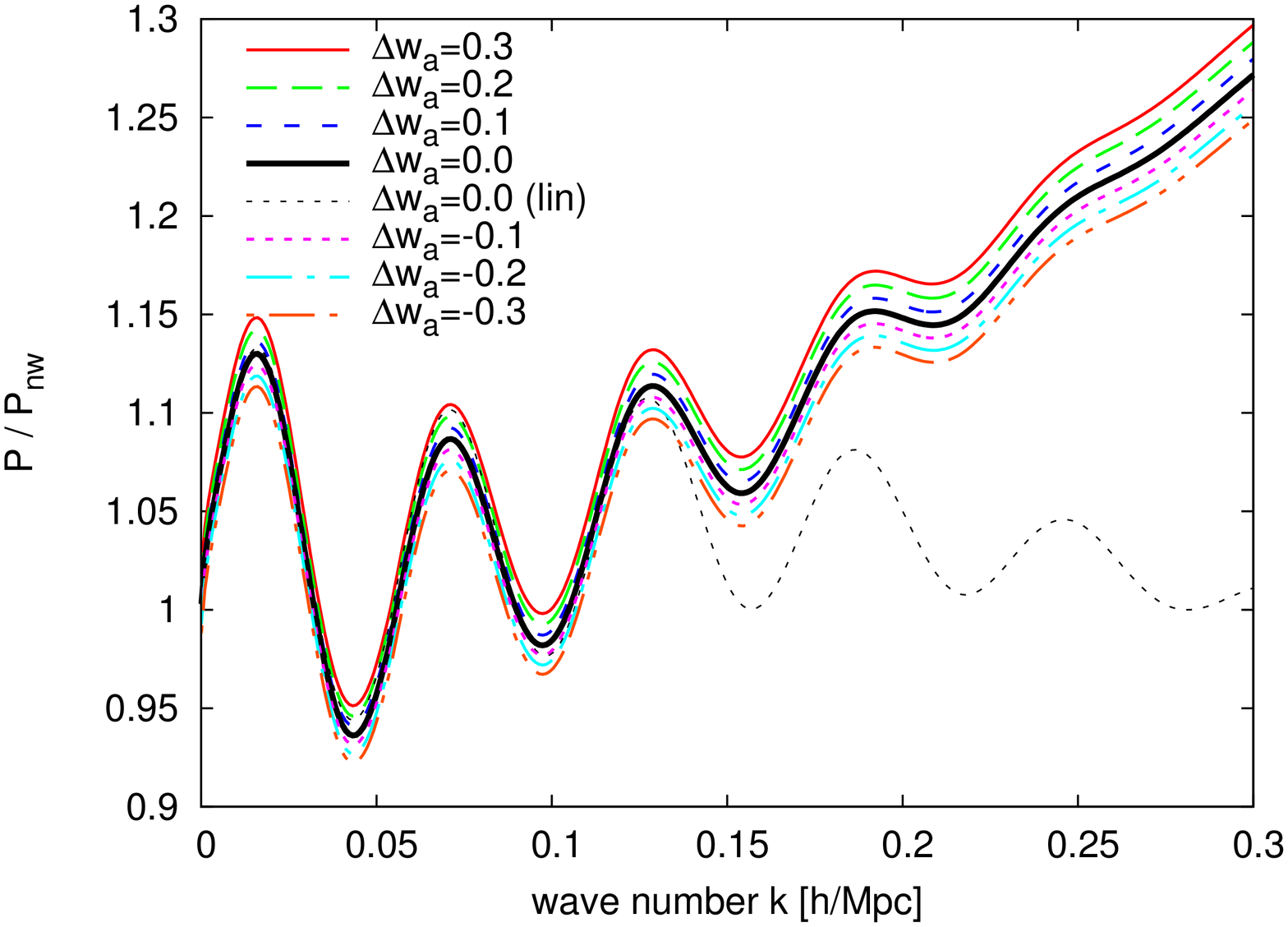}
\caption{Effects of varying $w_0$ (top) and $w_a$ (bottom) on the
  Time-RG matter power spectrum $P(k)$ at $z=1$.  The fiducial model
  $\Delta w_0$,~$\Delta w_a = 0$ is M000n0 in Table~\ref{tab:basic}.
  $P(k)$ has been divided by the no-wiggle power
  spectrum~(\ref{e:Pnw_r_r}) for clarity.\label{f:varyw_r}}
\end{center}
\end{figure}

Figure~\ref{f:varyw_r} shows the effects on the power spectrum of
varying the equation of state parameters $w_0$ and $w_a$, starting
from the $\Lambda$CDM fiducial model M000n0 from
Table~\ref{tab:basic}.  Power spectra are calculated using the Time-RG
perturbation theory; we have divided the power spectra by a smoothed
``no-wiggle'' power spectrum $\Pnw(k)$ specified in
Eq.~(\ref{e:Pnw_r_r}).

Our expectation based on linear theory, that changing the equation of
state mainly affects $P(k)$ through the growth factor, is essentially
correct for $w_0$.  The different curves in Fig.~\ref{f:varyw_r}~(top
panel) differ mostly by a normalization factor, corresponding to the
square of the growth factor at $z=1$.  On the other hand, changing
$w_a$ appears, from Fig.~\ref{f:varyw_r}~(bottom panel), to have a
greater effect at more nonlinear scales.  This is encouraging, as it
indicates that the nonlinear power spectrum can provide more powerful
constraints on $w_a$ than expected from linear theory.

\subsubsection{Massive Neutrinos}

\begin{figure}[tb]
\begin{center}
\includegraphics[angle=270,width=3.3in]{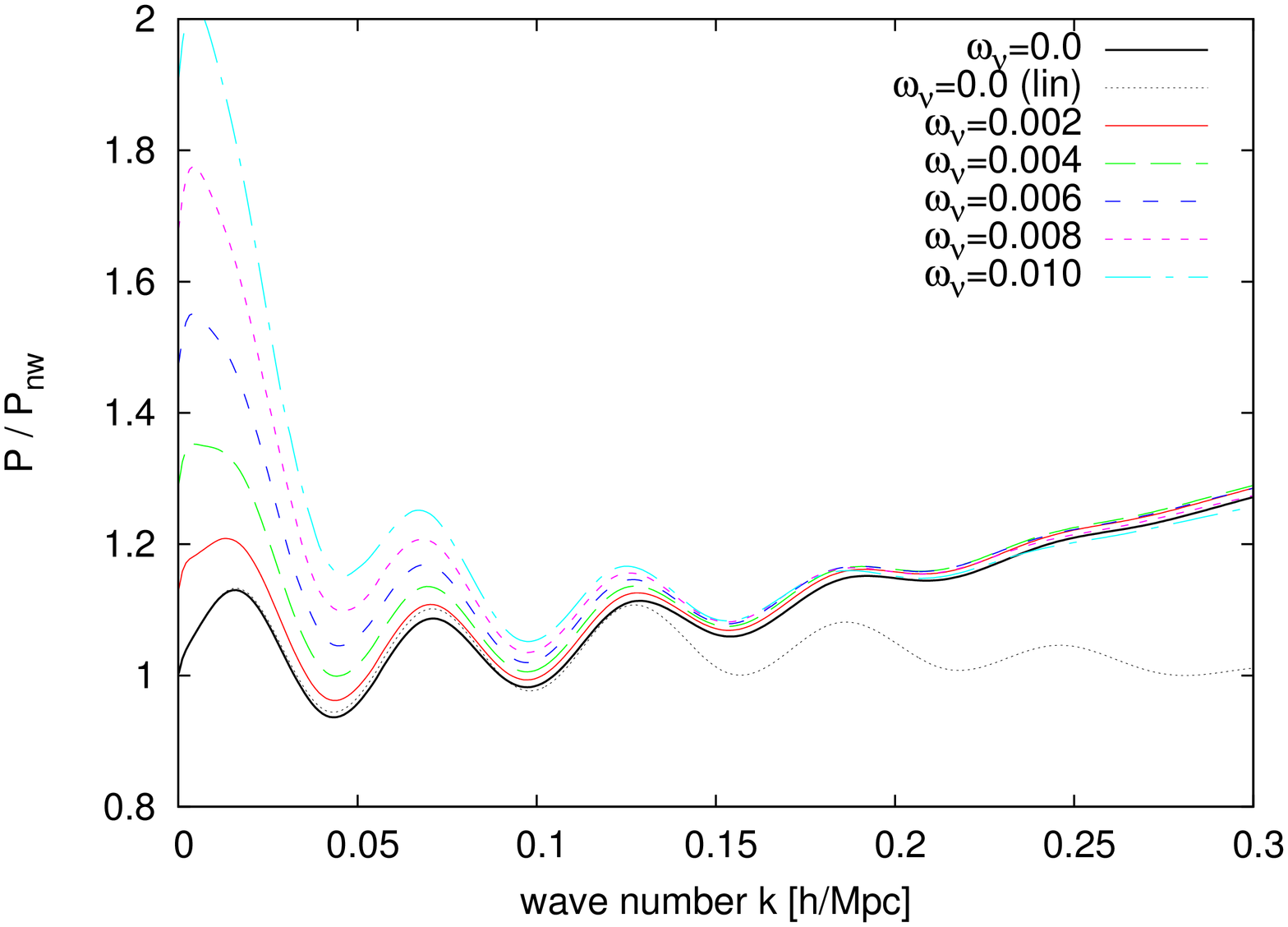}
\includegraphics[angle=270,width=3.3in]{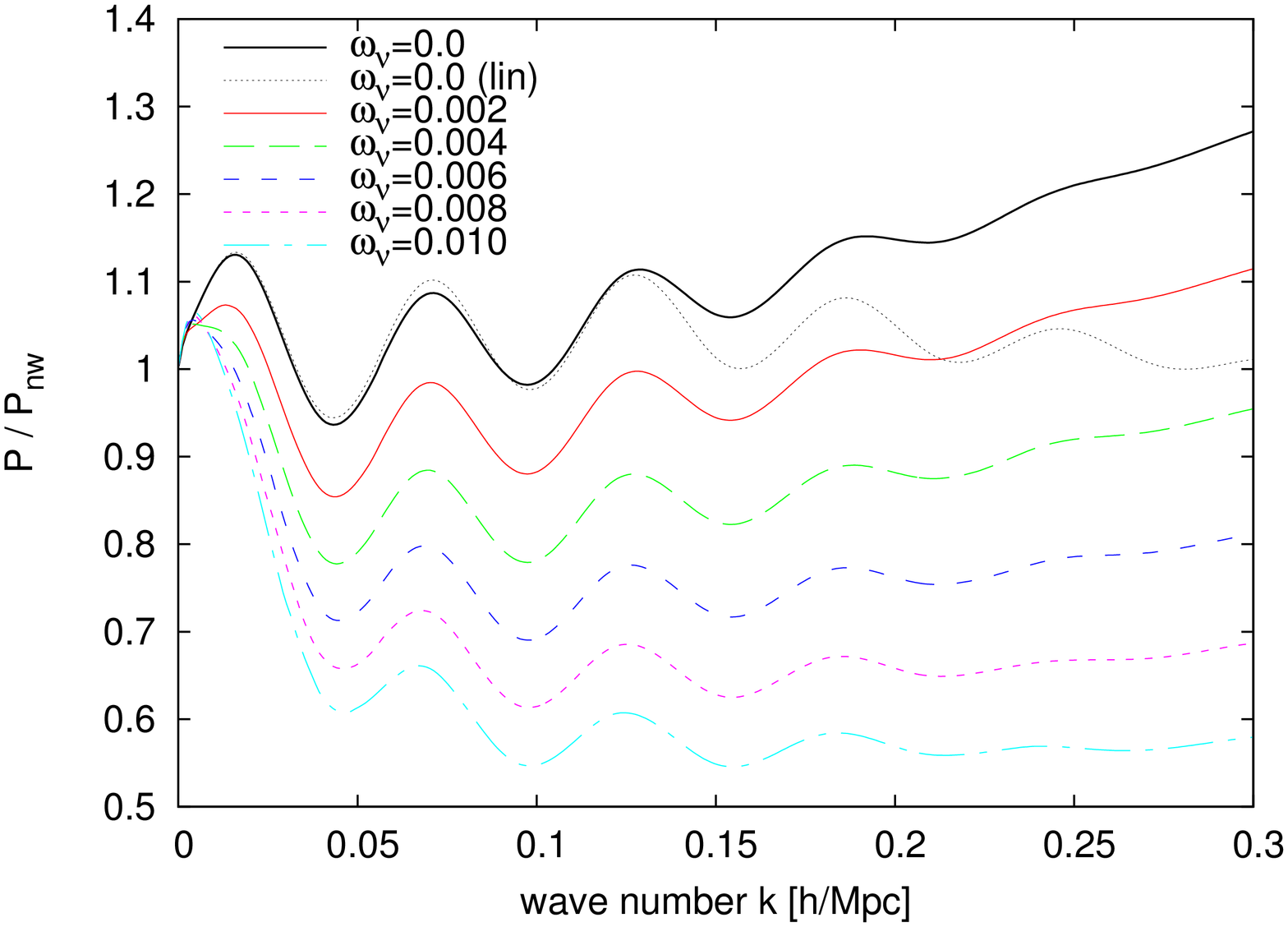}
\caption{Effects of varying $\omega_\nu$ on the Time-RG matter power
  spectrum at $z=1$. In each case the M000n0 values of
  $\omega_\mathrm{m}$, $n_s$, $h$, $w_0$, and $w_a$ are assumed. Top:
  $\sigma_8=0.8$ is fixed for all of the models. Bottom:~All models
  are normalized to the same low-$k$ value.  \label{f:vary_omnu}}
\end{center}
\end{figure}

\begin{figure}[tb]
\begin{center}
\includegraphics[angle=270,width=3.3in]{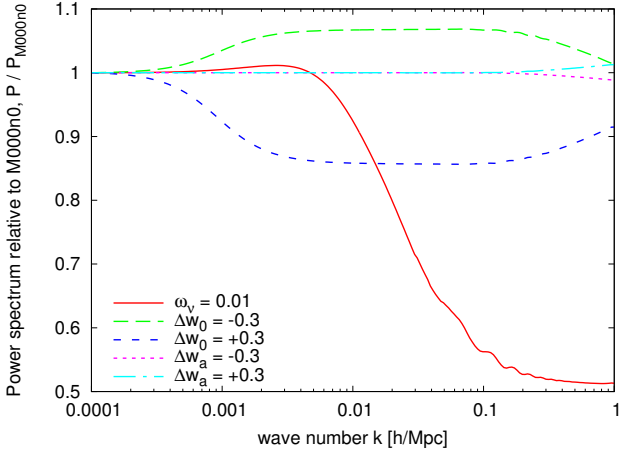}
\caption{
  Effects of varying $w_0$, $w_a$, and $\omega_\nu$  on the Time-RG matter
  power spectrum at $z=1$, relative to their M000n0 values.  All models
  are normalized to the same low-$k$ value.
  \label{f:ratio}
}
\end{center}
\end{figure}

\begin{table}
\begin{center}
  \caption{Wave number $k$ [$h$\mpcinvtxt] up to which linear and Time-RG
    perturbation theories are accurate up to $1\%$ (or $2\%$) in models
    with massive neutrinos.
    \label{t:PTerror_nu}}
\begin{tabular}{lcc|cc}
Model & $z$  & Acc. & Linear       & Time-RG \\
\hline
M000n1 & $0$ & $1\%$ &0.086 & 0.14 \\
       &     & \Tabg{$2\%$} & \Tabg{0.091} & \Tabg{0.14} \\
       & $1$ & $1\%$ &0.095 & 0.2 \\
       &     & \Tabg{$2\%$} & \Tabg{0.1} & \Tabg{0.26} \\

M000n2 & $0$ & $1\%$ &0.099 & 0.16 \\
       &     & \Tabg{$2\%$} & \Tabg{0.11} & \Tabg{0.2} \\
       & $1$ & $1\%$ &0.11 & 0.44 \\
       &     & \Tabg{$2\%$} & \Tabg{0.14} & \Tabg{0.58} \\

M001n1 & $0$ & $1\%$ &0.065 & 0.17 \\
       &     & \Tabg{$2\%$} & \Tabg{0.09} & \Tabg{0.17} \\
       & $1$ & $1\%$ &0.16 & 0.25 \\
       &     & \Tabg{$2\%$} & \Tabg{0.16} & \Tabg{0.26} \\

M002n1 & $0$ & $1\%$ &0.11 & 0.17 \\
       &     & \Tabg{$2\%$} & \Tabg{0.11} & \Tabg{0.17} \\
       & $1$ & $1\%$ &0.11 & 0.64 \\
       &     & \Tabg{$2\%$} & \Tabg{0.12} & \Tabg{0.78} \\

\end{tabular}
\end{center}
\end{table}
Figure~\ref{f:vary_omnu} (top panel) begins with model M000n0 and
increments $\omega_\nu$ in steps of $0.002$, with $\omega_\nu = 0.01$
corresponding to model M000n1.  As for the simulations, the total
matter power spectrum $P(k)$ is found by adding the nonlinear CDM and
baryon power spectrum to the linear neutrino power spectrum (see
Eq.~\ref{e:pk}).  Increasing the neutrino mass modifies the power
spectrum in a scale-dependent way.  We note that all of the models in
Fig.~\ref{f:vary_omnu} (top panel) are normalized to the same
$\sigma_8$ at $z=0$.  Since neutrinos
suppress small-scale power, normalization increases the large-scale
power of massive neutrino models to compensate.  The bottom panel of
Fig.~\ref{f:vary_omnu} shows the results if one fixes the
normalization by adopting the same low-$k$ amplitude in all cases, at
some chosen value of $k$. In this case, the suppression of power due
to massive neutrinos is immediately evident.

\begin{figure}[tb]
\begin{center}
\includegraphics[angle=270,width=3.3in]{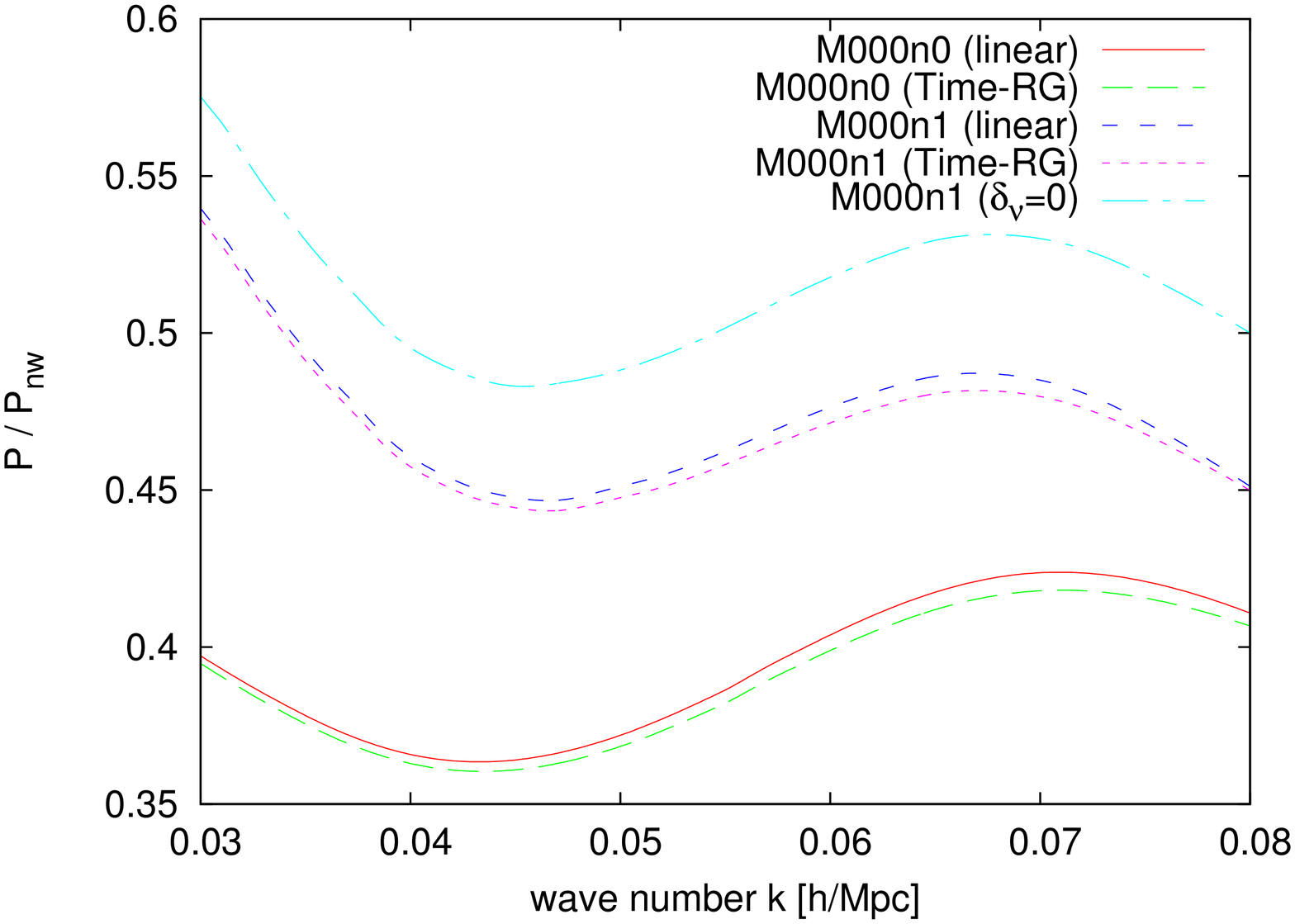}
\includegraphics[angle=270,width=3.3in]{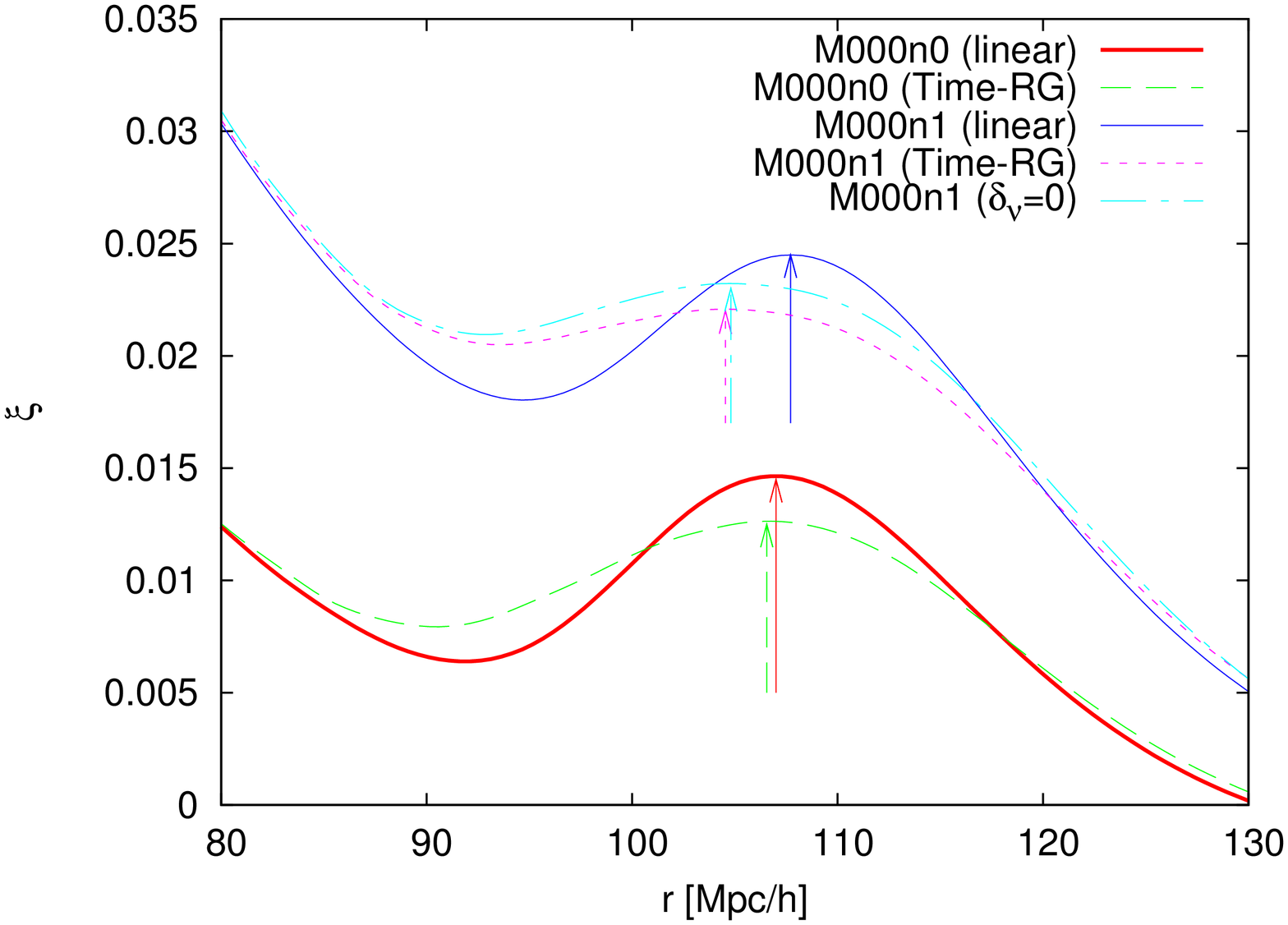}
\caption{Top panel: Shifts in the baryon wiggles in the matter power
  spectrum at $z=1$ due to the effects of neutrino mass.
  Lower panel: Correlation function at $z=1$.  
  Vertical arrows show the locations of local extrema. 
  The shift is almost entirely due to the neutrino 
  contribution to the homogeneous expansion rate
  $H(z)$.\label{f:neutrino_shifts}}
\end{center}
\end{figure}

Massive neutrinos shift the positions of the baryon peaks in the power
spectrum, with possible implications for BAO measurements.  For
example, the $k$ positions of the trough at $k\approx 0.04~h\mpcinv$
and the peak at $k \approx 0.07~h\mpcinv$ are shifted relative to the
massless neutrino model in Fig.~\ref{f:neutrino_shifts}.  Comparing
the $\omega_\nu=0$ (red, solid) and $\omega_\nu=0.01$ Time-RG (green,
long-dashed) curves in that figure, we see that neutrino masses shift
the $k$ values of these extrema by $\sim \fnu$.
Figure~\ref{f:neutrino_shifts} also shows the spatial correlation
function $\xi(r)$ of the matter, computed using the {\tt FFTLOG}
package~\cite{Hamilton00}. Transformation from $P(k)$ to $\xi(r)$
requires the extrapolation of $P$ to large $k$, which we do using a
power law of slope $d\log P / d \log k = n_s - 3$ for linear $P(k)$
and $-1$ for nonlinear $P(k)$.  Varying this slope by $\pm 30\%$
changes the BAO peak position by only $0.1\%$, so the BAO feature is
robust with respect to this extrapolation.


The neutrino contribution to the power spectrum can be divided into
two effects.  The first is the effect of the neutrino energy density
on the homogeneous expansion $H(z)$ of the universe, including the
resulting effect on the CDM and baryon growth factor.  The second is
the direct contribution of the neutrino inhomogeneities $\delta_\nu
\neq 0$ to the total matter power spectrum and to the scale-dependent
growth of $\delcb$.  In order to separate these effects,
Fig.~\ref{f:neutrino_shifts} also shows the $\omega_\nu=0.01$ power
spectrum with the neutrino inhomogeneities set to zero,
$\delta_{\nu\mathrm{,lin}} = 0$ in Eq.~(\ref{e:Omega10_nu}).  Evidently
the shifts in the baryon wiggles in the CDM power spectrum are almost
entirely due to the first effect, the neutrino contribution to $H(z)$;
the peak and trough positions in the Time-RG and $\delta_\nu=0$ curves
differ by $\sim 0.1\%$.  

In terms of BAO analysis, this is a potentially helpful result. It
implies that BAO reconstruction, which uses 2-loop Lagrangian
perturbation theory to map the observed density field back to the
underlying linear field, can be trivially extended to include massive
neutrinos.  In order to do this, one need only include the energy
density and pressure of the neutrinos when computing $H(z)$ and the
(scale-independent) CDM+baryon growth factor. If one approximates the
BAO peak by ignoring neutrino inhomogeneities in Time-RG, then the
peak position is off by less than $0.25\%$. This level of
approximation is more than adequate for BOSS, but may become important
under optimistic assumptions about DESI ~\cite{font-ribera13}.

\section{Conclusion}
\label{sec:conclusion}

Over the next several years, surveys measuring the BAO peak and the
growth of large-scale structure will provide substantially improved
constraints on dynamical dark energy and massive neutrinos, especially
when combined with lensing observations~\cite{font-ribera13}, as well
as with CMB measurements.  In particular, they will significantly
narrow the allowed range of dark energy equations of state and will
measure, rather than merely bound, the sum of neutrino masses.
Analyses of these upcoming data will require a thorough understanding
of the subtle effects on the matter power spectrum arising from the
dark energy and neutrino sectors.

Higher-order perturbation theory and N-body simulations provide
complementary predictions of the matter power spectrum, overlapping at
quasilinear scales.  In this work we have extended both tools to
cosmologies with time-varing dark energy equations of state and
massive neutrinos.  By modifying the publicly available
\copter~code~\cite{COPTER}, we extended several higher-order
perturbation theories to cosmologies with arbitrary homogeneous
evolution $H(z)$.  Figure~\ref{f:allPT_r} and Table~\ref{t:PTerror}
compare linear theory and six different higher-order perturbation
theories to N-body simulations for a $\Lambda$CDM model as well as an
early dark energy.  For $\Lambda$CDM (model M000n0) the higher-order
calculations all agree with simulations to $2\%$ up to $k=0.1$~h\mpcinvtxt
at $z=0$ and up to $k=0.15$~h\mpcinvtxt at $z=1$; some of the perturbation
theories perform substantially better than that.  For early dark energy
(model M001n0) the situation is similar except that $2$-loop SPT
behaves badly at $z=0$.

In addition to arbitrary homogeneous evolution, we included massive
neutrinos, treated linearly, in the Time-RG perturbation theory.  Our
results in Figs.~\ref{f:varyw_r}-\ref{f:vary_omnu} show the effects of
incrementing $w_0$, $w_a$, and $\omega_\nu$ on the matter power
spectrum.  We find the interesting result that, on BAO scales, the
neutrino contribution to the CDM and baryon power spectrum is
dominated by the neutrinos' modification to the homogeneous expansion
rate $H(z)$, as shown in Fig.~\ref{f:neutrino_shifts}.  Neglecting
neutrino inhomogeneities in the standard Lagrangian Resummation
Perturbation Theory reconstruction will therefore only introduce an
error of $\leq 0.25\%$ in the position of the BAO peak, assuming a
neutrino-to-matter ratio $f_\nu \leq 0.075$.

We added neutrinos to the HACC N-body code in a minimal fashion by
neglecting the neutrino density contrast as a source for matter
clustering. The matter power spectrum is found by combining the
nonlinear CDM plus baryon power spectrum result along with that from
neutrinos treated in linear theory. This approximation can be tested
directly within Time-RG perturbation theory by neglecting the neutrino
density contrast in Eq.~(\ref{e:Omega10_nu}).  For $f_\nu=0.075$, at
$z=0$, we find that the approximation is valid to better than $1\%$ up
to $k=0.16~h$\mpcinvtxt, essentially the entire range of validity of
perturbation theory.  Since $f_\nu=0.075$ is a few times larger than
allowed by current constraints, we are justified in applying this
approximation for N-body calculations.  Figures~\ref{f:m000n}
and~\ref{f:nuEDE_r} compare the resulting N-body power spectra to
Time-RG with massive neutrinos in $\Lambda$CDM and early dark energy
models, respectively.  Table~\ref{t:PTerror_nu} summarizes the
results; at $z=0$, Time-RG and simulations agree to $2\%$ up to at least
$k=0.14~h$\mpcinvtxt for all models considered.  The combination of
our Time-RG and N-body calculations is a powerful result, predicting
the power spectrum in massive neutrino models, for a wide range of
dark energy models, over several orders of magnitude in $k$.

\subsection*{Acknowledgments}
We are grateful to J.~Kwan, Z.~Luki\'c, and M.~Pietroni for insightful
conversations.  SH thanks Masahiro Takada for useful discussions. The
authors were supported by the U.S. Department of Energy, Basic Energy
Sciences, Office of Science, under contract No. DE-AC02-06CH11357.
This research used resources of the ALCF, which is supported by DOE/SC
under contract DE-AC02-06CH11357 and resources of the OLCF, which is
supported by DOE/SC under contract DE-AC05-00OR22725.

The submitted manuscript has been created by UChicago Argonne, LLC,
Operator of Argonne National Laboratory (``Argonne''). Argonne, a
U.S. Department of Energy Office of Science laboratory, is operated
under Contract No. DE-AC02- 06CH11357. The U.S. Government retains for
itself, and others acting on its behalf, a paid-up nonexclusive,
irrevocable worldwide license in said article to reproduce, prepare
derivative works, distribute copies to the public, and perform
publicly and display publicly, by or on behalf of the Government.


\end{document}